\documentclass[fleqn,usenatbib]{mnras}

\usepackage{newtxtext,newtxmath}
\usepackage[T1]{fontenc}

\DeclareRobustCommand{\VAN}[3]{#2}
\let\VANthebibliography\thebibliography
\def\thebibliography{\DeclareRobustCommand{\VAN}[3]{##3}\VANthebibliography}

\usepackage{subfig}     % Format sub-figures (a, b, etc)

\usepackage{graphicx}	% Including figure files
\usepackage{amsmath}	% Advanced maths commands

\defcitealias{Patton2020}{P2020}

\title[Interacting Galaxies in IllustrisTNG -- V]{Interacting galaxies in the IllustrisTNG simulations -- V: Comparing the influence of star-forming vs. passive companions}

\author[W. Brown et al.]{
Westley Brown$^{1,2}$,\thanks{E-mail: westleyb@yorku.ca}
David R. Patton$^{1}$,
Sara L. Ellison$^{3}$,
Lawrence Faria$^{1,4}$
\\
$^{1}$Department of Physics and Astronomy, Trent University, 1600 West Bank Drive, Peterborough, ON K9L 0G2, Canada\\
$^{2}$Department of Physics and Astronomy, York University, 4700 Keele Street, Toronto, ON M3J 1P3, Canada\\
$^{3}$Department of Physics and Astronomy, University of Victoria, Finnerty Road, Victoria, British Columbia, V8P 1A1, Canada\\
$^{4}$Department of Physics, Engineering Physics and Astronomy, Queen's University, 64 Bader Lane, Kingston, ON K7L 3N6, Canada\\
}

\date{Accepted XXX. Received YYY; in original form ZZZ}

\pubyear{2020}

\begin{document}
\label{firstpage}
\pagerange{\pageref{firstpage}--\pageref{lastpage}}
\maketitle

\begin{abstract}
We study interacting galaxy pairs in the TNG100-1 and TNG300-1 cosmological simulations using previously generated closest companion samples. We study the specific star formation rates (sSFR) of massive ($10^{10} M_{\odot} < M_* < 10^{12} M_{\odot}$) galaxies at $z \leq 0.2$ as a function of separation from the closest companion galaxy. We split our sample based on whether the companion galaxy is star-forming or passive. We find that galaxies with close \textit{star-forming} companions have sSFRs that are enhanced (on average) by a factor of $2.9 \pm 0.3$ in TNG100-1 and $2.27 \pm 0.06$ in TNG300-1 compared to controls,  with enhancements present out to separations of $\sim 300$ kpc. Galaxies with \textit{passive} companions in TNG300-1 exhibit mild sSFR suppression ($\sim12$ percent) at 100-300 kpc and small sSFR enhancements at separations below 50 kpc. sSFR suppression is strongest in pairs where the galaxy’s stellar mass is more than 2 times that of its passive companion. By generating a stellar mass-matched ("twinned") sample in TNG300-1, we show that differences in sSFR trends between companion types are not a result of intrinsic stellar mass differences in star-forming vs. passive galaxies. We compare with an analogous sample of galaxy pairs from SDSS, finding consistent results between observations and simulations. Overall, we find that star-forming galaxies show enhanced sSFRs regardless of companion type, but that galaxies with close passive companions are more likely to be passive themselves.
\end{abstract}


%%%%%%%%%%%%%%%%%%%%%%%%%%%%%%%%%%%%%%%%%%%%%%%%%%

%%%%%%%%%%%%%%%%% BODY OF PAPER %%%%%%%%%%%%%%%%%%

\section{Introduction}
\label{section:intro}

Merging galaxies have fascinated astronomers for decades, as they paint an elegant and complex picture of galaxies in motion, driven by strong gravitational attractions. While we can only observe a single point in time for any physical galaxy pair, the long timescales involved in galaxy mergers allow us to study each stage of the merging process in detail across multiple pairs. Even in the long interaction stage before a merger, galaxies exhibit distinct visual features such as tidal tails and morphological distortions \citep{dePropris2007,Lotz2008,Casteels2014, Patton2016}, as well as diluted metallicities \citep{Kewley2006,Ellison2008,Scudder2012,Torrey2012,Bustamante2018,Bustamante2020}, higher fractions of active galactic nuclei (AGN) \citep{Ellison2011,Satyapal2014,Ellison2019,Mcalpine2020}, and enhanced star formation rates (SFR) \citep{Ellison2008,Scudder2012,Patton2013,Cao2016,Patton2020} compared to galaxies not involved in interactions.

Like many of the phenomena associated with galaxy interactions, studies have repeatedly found strong anti-correlations between SFR enhancements and galaxy separations, such that galaxy pairs with the smallest separations typically have the largest enhancements. In addition, SFR enhancements have been found out to projected separations as large as $\sim 150$ kpc \citep{Patton2013}, suggesting that lower level enhancements in wider pairs were triggered by recent pericentre passages rather than ongoing interactions. Simulations of merging galaxy pairs confirm this interpretation, exhibiting SFR enhancements that are triggered during close encounters \citep{Renaud2022}, and an anti-correlation between SFR enhancements and pair separation that arises from the increasing separation of galaxies following close encounters \citep{Scudder2012, Patton2013, Moreno2015}.  Merger simulations provide additional insights into the role that gravitational torques and hydrodynamical processes play during these interactions \citep{Torrey2012,Hopkins2013,Moreno2019,Moreno2021,Sparre2022}.

The gravitational picture of star formation is a relatively simple one, which tells us that star formation occurs when gas reaches the critical density required to self-collapse. In this picture, the main drivers of SFR enhancement in interacting galaxies are mass and separation. As we add in more detail, we begin to understand that only a specific kind of gas can form stars---cold  molecular hydrogen (\textsc{H}$_2$) gas. The forces involved in typical galaxy interactions lead to an inflow of cold gas to the centres of galaxies, prompting star formation in these areas as seen in both observations \citep{Barrera-Ballesteros2015,Chown2019,Pan2019} and simulations \citep{Hopkins2013,Blumenthal2018,Moreno2019,Moreno2021}.

Some studies of SFR enhancement in galaxy pairs have tried to look beyond simple gravitational effects, and introduced additional variables such as galaxy morphology and star formation classification. For example, \citet{Xu2010} and \citet{Cao2016} found that spiral galaxies interacting with other spiral galaxies typically have enhanced SFRs, while spiral galaxies interacting with elliptical galaxies do not. \citet{Moon2019} conducted a study of galaxy pairs at $0.01 < z < 0.1$ from the Sloan Digital Sky Survey (SDSS), which they then split into four categories based on whether each galaxy in the pair was star-forming or quiescent. They found that while galaxies with close star-forming companions have enhanced SFRs compared to control galaxies without companions, galaxies with close quiescent companions instead showed evidence of suppressed SFRs compared to controls of the same type.

In a study of interacting SDSS galaxies, \citet{Park2009} suggest that the strength of interaction-driven star formation effects as a function of virial radius provides “strong evidence for hydrodynamical interactions”. Indeed, the results of \citet{Xu2010}, \citet{Cao2016}, and \citet{Moon2019}  which examine interactions of different types of galaxies, all suggest that gravitational forces alone are not sufficient to explain enhanced SFRs in galaxy interactions, and suggest that hydrodynamical interactions may play a significant role. If all masses and separations were equal, gravity alone would be insufficient to explain differences in SFR enhancement between galaxy types.

\citet{Jog1992} initially proposed one hydrodynamical explanation: that collisions between the interstellar medium (ISM) of both pair galaxies may be an important hydrodynamical component to SFR enhancement in galaxy interactions. \citet{Barnes2004} and \citet{Braine2004} furthered this theory by investigating the impact of shock-induced star formation caused by ISM collisions. In a recent study of simulated galaxies, \citet{Renaud2022} find that tidal compressions from close galaxy interactions and mergers trigger starbursts in galaxies with galactic discs. Conversely, hydrodynamics may also explain possible SFR suppression in some interacting galaxies. \citet{Correa2018} discuss how some massive galaxies may build up a large halo of hot gas, and if these halos were to overlap with the gas halo of another galaxy they may be sufficient to effectively cut off the flow of cold gas into the other galaxy, reducing its ability to replenish gas to form new stars.

Since we cannot directly observe the changes to flow or temperature of gas within real galaxies on measurable timescales, the best option we have to study these phenomena and assess these theories is through simulations. Assuming that a simulation is a decent representation of real physics, it can provide us with more detail than observations and give us a key to understanding what we see in the real universe. Simulations also provide other advantages over observations, such as 3D positional information, which allows one to avoid the projection effects that observers must contend with.  High-resolution simulations have long been used in the study of gas flows within interacting galaxies; however, these simulations often focus on idealized interactions and mergers between two star-forming galaxies \citep{Cox2008,diMatteo2007,Lotz2008,Torrey2012}, neglecting the role of additional companions \citep{Moreno2013}.

Cosmological simulations such as Illustris \citep{Vogelsberger2014}, EAGLE \citep{Schaye2015}, SIMBA \citep{Dave2019} and IllustrisTNG \citep{Nelson2019a} offer large samples of galaxy pairs evolved in a wide variety of realistic galaxy environments, at the expense of lower resolution and longer gaps in time sampling compared to merger simulations or so-called "zoom-in" simulations. Even with lower resolution, IllustrisTNG’s first two flagship cosmological simulations have proven to be well-suited to the study of SFR enhancements both in general \citep{Torrey2018,Donnari2019,Piotrowska2022} and in interacting and merging galaxies \citep{Patton2020,Hani2020,Quai2021}.

Previous papers in this series all use IllustrisTNG to study the properties of interacting and merging galaxies. \citet{Patton2020} study enhanced sSFR in interacting galaxies as a function of pair separation. They also compare their IllustrisTNG samples with an analogous sample of SDSS galaxy pairs, and find similar sSFR enhancements in IllustrisTNG and SDSS. \citet{Hani2020} conduct a similar study of sSFR trends in post-merger galaxies in IllustrisTNG, finding that star-forming post-mergers on average have enhanced sSFR compared to controls while passive post-mergers have similar sSFR to controls. \citet{Quai2021} examine the quenching rates and timescales of post-merger galaxies and find that while quenched post-mergers are rare, quenching occurs at twice the rate of control galaxies within 500 Myr after coalescence. Meanwhile, \citet{Byrne-Mamahit2023} look at supermassive black hole accretion rates in post-merger galaxies, finding that post-mergers have higher accretion rates than controls, and that these enhanced accretion rates persist longer than post-merger SFR enhancement.

To further the investigation of \citet{Patton2020}’s study (hereafter, \citetalias{Patton2020}), this paper will use the same sample of IllustrisTNG galaxy pairs to analyze whether differences in companion type lead to differences in the trend of SFR enhancements within the simulation. Companion types will be categorized as either star-forming or quiescent/passive, similar to \citet{Moon2019}. Using a series of analyses, we seek to determine the differences in sSFR trends with companion type and explore possible explanations behind those differences.

The paper is organized as follows. Section \ref{section:methods} describes the main methods used in the generation of the sample, including the IllustrisTNG simulations and the measurement of SFR enhancements. Section \ref{section:resultsMain} covers the results of the separation by companion type, including a breakdown by stellar mass ratio. In Section \ref{section:mass-analysis}, we aim to eliminate mass differences by generating “twinned” samples of galaxies with star-forming and passive companions. In Section \ref{section:observational}, we repeat our analysis in projected space and compare our simulation results with observational data from SDSS. In Section \ref{section:host_type}, we provide an additional layer of analysis by separating our sample into star-forming and passive host galaxies. In Section \ref{section:discussion} and Section \ref{section:conclusions} we present our discussion and conclusions.

%%%%%%%%%%%%%%%%%%%%%%%%%%%%%%%%%%%%%%%%%%%%%%%%%%

\section{Methods}
\label{section:methods}

\subsection{TNG100-1 and TNG300-1 simulations}
\label{section:sims}

This paper will mainly utilize data from IllustrisTNG, a suite of cosmological gravo-magnetohydrodynamical simulations which model the evolution of dark matter, gas, stars, and supermassive black holes using the moving-mesh code \textsc{Arepo} \citep{Springel2010}. IllustrisTNG is the successor to the Illustris simulations \citep{Vogelsberger2014}, and provides a number of improvements to the original simulation suite. The full data release can be found in \citet{Nelson2019a}, and additional details including stellar content of groups and clusters, galaxy clustering, galaxy colour bimodality, chemical enrichment, and magnetic fields can be found in \citet{Pillepich2018b}, \citet{Springel2018}, \citet{Nelson2018a}, \citet{Naiman2018}, and \citet{Marinacci2018}.

Our analysis focuses on the highest resolution runs of TNG100 and TNG300. TNG100-1 is run on a three-dimensional cube of volume (110.7 Mpc)$^3$ with a baryonic mass resolution of $1.36 \times 10^6$ M$_{\odot}$. TNG300-1 covers a much larger volume of (302.6 Mpc)$^3$ at the expense of a lower baryonic mass resolution of $1.1 \times 10^7$ M$_{\odot}$, which is roughly equivalent to the resolution of the TNG100-2 run. Each simulation is released in 100 "snapshots" corresponding to different ages of the universe, beginning at snapshot 0 (redshift $z = 127$) and ending at snapshot 99 (redshift $z=0$). The average time between snapshots is approximately 160 Myr. Star formation occurs stochastically within the simulations when gas particles exceed a star formation density threshold, following the Kennicutt-Schmidt relation and assuming a Chabrier initial mass function \citep{Pillepich2018a}. Star formation rates are computed by summing the instantaneous SFR of all gas particles within a specified volume or assigned to a particular subhalo.

The \textsc{Subfind} algorithm \citep{Springel2001,Dolag2009} has been run on each snapshot to identify gravitationally-bound subhaloes (galaxies) and designate the gas, stellar, and dark matter particles correspondingly. As discussed in-depth in \citet{Rodriguez-Gomez2015} and \citetalias{Patton2020}, this can sometimes lead to crowding issues (also called numerical stripping) when two or more subhaloes are in close enough proximity such that particles from one subhalo can be mistakenly assigned to the other. For that reason, one must be careful when studying close galaxy pairs within these simulations. As with \citetalias{Patton2020}, when computing specific SFR we choose to take stellar masses and SFR values from within one stellar half-mass radius ($R_{1/2}$) to minimize contamination from numerical stripping. This is acceptable for our purposes, as it has been shown that SFR enhancements between interacting galaxies tends to be concentrated within the centres of galaxies in both simulations \citep{Moreno2015,Moreno2021} and observations \citep{Ellison2013,Chown2019,Pan2019,Thorp2019,Thorp2022}.

\subsection{Samples of galaxy pairs from P2020}
\label{section:sample}

We utilize two samples of closest companion galaxies in IllustrisTNG identified by \citetalias{Patton2020}, which allows us to compare the higher resolution of the TNG100-1 simulation with the significantly larger sample size of the TNG300-1 simulation. The original samples feature pairs of interacting galaxies from snapshot 50 through 99 (redshift $0\leq z \leq 1$), where each pair consists of one "host" galaxy and its closest "companion" galaxy. Any galaxy in IllustrisTNG may be chosen as a host galaxy once, but may also appear in the sample as the closest companion to another host galaxy multiple times. In the majority of pairs, host galaxies are more massive than their companions, but this is not a requirement.

In developing these samples, \citetalias{Patton2020} impose stellar mass limits on all host galaxies of $10^{10} M_{\odot} < M_* < 10^{12} M_{\odot}$, where $M_*$ is the best estimate of a galaxy's total stellar mass. For most galaxies, $M_*$ is the current stellar mass reported in the IllustrisTNG database.  However, if the 3D separation between the host and companion is less than twice the sum of their stellar half-mass radii, the best estimate of the stellar mass is taken to be the maximum stellar mass of the galaxy within the last 0.5 Gyr. \citetalias{Patton2020} demonstrate that this approach is effective at minimizing the effects of numerical stripping in close overlapping pairs.

Each companion galaxy is defined as the subhalo with the smallest 3D separation ($r$) to its host, and must have a total stellar mass of at least 0.1 times that of the host. This results in a minimum stellar mass limit of $10^9 M_{\odot}$, which is sufficient to ensure that all relevant companions are well resolved \citepalias{Patton2020}. Galaxy pairs with companions that have a stellar mass of more than 10 times that of the host are later removed from the sample. Throughout this paper, $r$ is defined as the 3D distance between the centre of the host galaxy and the centre of its companion galaxy.

Each host galaxy is also matched to exactly one control galaxy. The aim of control matching is to find a single control galaxy which best resembles the host galaxy's stellar mass, environment, and redshift without the additional influence of the closest companion galaxy. In the \citetalias{Patton2020} samples, control galaxies are matched to each host by exactly matching the snapshot (equivalent to redshift), and by closely matching the parameters of stellar mass ($M_*$), number of companions within 2 Mpc ($N_2$), and distance to the second-closest companion ($r_2$). The methods used to match these control parameters and determine the best match are detailed in \citet{Patton2016} and \citetalias{Patton2020}. We direct curious readers to Figure 4 of \citetalias{Patton2020}, which shows how control-matching parameters such as mass and environment scale with pair separation, $r$. We note that many of our plots also include galaxies with closest companions found at extreme separations as large as 1 Mpc, and that galaxy pairs such as these reside in low-density environments, such as cosmic voids. The environmental parameters included in the control matching process ensure that these pairs are each matched with a control galaxy in a similar low-density environment.

\citetalias{Patton2020} also do extensive work to determine the limits of crowding and numerical stripping within their samples. Through a process of comparing the 3D galaxy separations to the overlap in $R_{1/2}$ of each galaxy pair numerically and through visual analysis of synthetic stellar composite galaxy images, \citetalias{Patton2020} estimate limits in $r$ for the relative stability of their subhalo data (hereafter referred to as crowding limits). For their TNG100-1 sample, they calculate a crowding limit of $r=15.9$ kpc, while in TNG300-1 they calculate a crowding limit of $r=17.4$ kpc. Caution is taken when analysing results below these limits as the sample becomes relatively incomplete and at risk of particle contamination and/or numerical stripping.

\subsection{Measuring sSFR enhancement}
\label{section:methodsSFR}

Throughout this paper, we define the variable $Q$ as the average sSFR enhancement (as in \citetalias{Patton2020} and \citealt{Hani2020}). This quantity is calculated using the averages in each $r$ bin as follows:
\begin{equation}
	Q\rm{(sSFR)} = \frac{\langle\rm{sSFR_{hosts}}\rangle}{\langle\rm{sSFR_{controls}}\rangle}.
\end{equation}

Since $Q$(sSFR) is a ratio, values above 1 indicate an enhancement of sSFR, while values between 0 and 1 indicate a suppression.

\begin{figure}
	\begin{center}
		\includegraphics[width=\columnwidth]{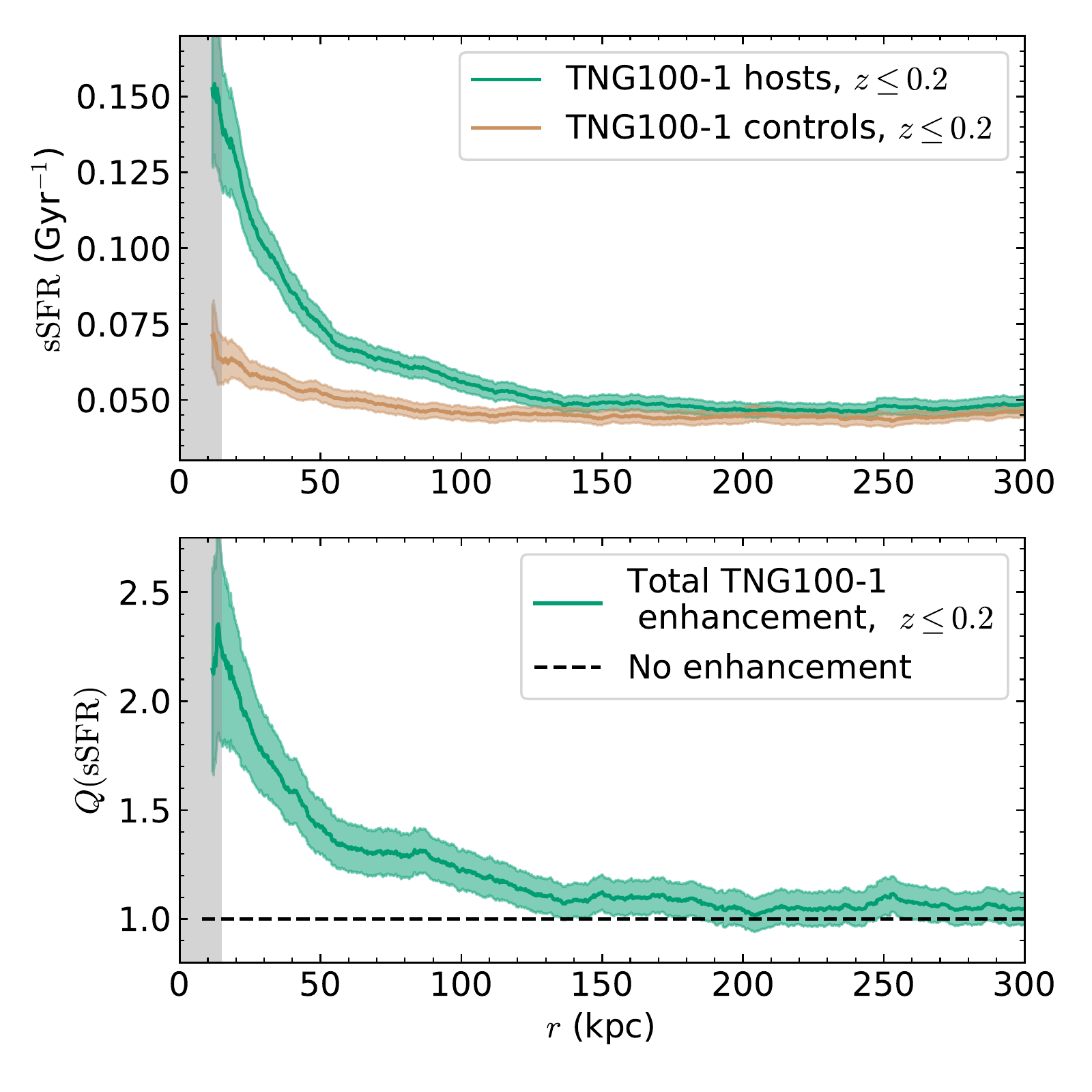}
		\caption{In the upper panel, the mean sSFR of all TNG100-1 host galaxies (green) and controls (tan) at $z \leq 0.2$ is plotted vs. 3D separation to the host's closest companion galaxy ($r$). In the lower panel, the sSFR enhancement ($Q$) is plotted vs. $r$, along with a black dashed line along $Q = 1$ which represents no enhancement. In both panels, the grey region represents an approximation of the crowding regime. Shaded regions in both panels represent the $2\sigma$ standard error in the mean.}
		\label{fig:West Master}
	\end{center}
\end{figure}

In many of our figures, we generate two plots, including sSFR vs. $r$ for both the hosts and controls as well as $Q\rm(sSFR)$ vs. $r$. This is exemplified in Fig. \ref{fig:West Master}, where we plot our unseparated, redshift-reduced sample (as discussed in \S{\ref{section:samplesep}}) of galaxy pairs in TNG100-1. All such figures have been made using a variable width box kernel, where the solid coloured lines indicate an average and the coloured shaded regions around them indicate an uncertainty of $2\sigma$. The shaded grey area indicates the approximate region below the crowding limit appropriate to the simulation, where the sample may be relatively incomplete and/or contaminated by numerical stripping, while the horizontal black dashed line at $Q=1$ in the bottom panel indicates no net enhancement.

As \citetalias{Patton2020}'s crowding limits are calculated specifically using their full sample, we choose to include demarcations of approximate estimates of these limits and caution the reader in interpreting results below these limits. Going forward, we place our crowding limit estimates at $r=15$ kpc in TNG100-1 and $r=17$ kpc in TNG300-1. As shown in Fig. \ref{fig:West Master}, our unseparated sample of $z \leq 0.2$ pairs has a maximum sSFR enhancement of $Q = 2.3 \pm 0.3$ at the crowding limit estimate of 15 kpc, which gradually decreases as separation increases. Enhancements are seen out to maximum separations of $r \sim 150$ kpc. This is in good agreement with \citetalias{Patton2020}'s full $z < 1$ sample.

\subsection{Separation of samples by companion type}
\label{section:samplesep}

We begin our analysis by restricting both the \citetalias{Patton2020} TNG100-1 and TNG300-1 samples to galaxies from snapshot 84 through 99 ($0 \leq z \leq 0.2$), to allow us to more easily compare with observational data and minimize evolution with redshift. Following \citetalias{Patton2020}, we treat each snapshot as a unique set of galaxies at a specific redshift, effectively increasing our sample size despite the limited volumes of the simulations. We then separate the samples into two categories each: host galaxies with star-forming companions (\textit{Hs}) and host galaxies with passive companions (\textit{Hp}). We use a threshold of sSFR = 0.01 Gyr$^{-1}$ to delineate between passive and star-forming galaxies, as in \citetalias{Patton2020} and \citet{Donnari2019}. After separating the sample, we assess each category to ensure a sufficient sample size for reliable analysis.

As part of our aim is to determine whether any differences in SFR effects between companion types are possibly hydrodynamical in nature, we also perform extensive mass analysis on our samples in order to determine the extent of mass differences between \textit{Hs} and \textit{Hp} pairs and the role that mass distribution plays in our sample. To that end, we further separate our samples into sub-categories based on their relative mass ratios. We define the stellar mass ratio of a host galaxy as:
\begin{equation}
	\mu = \frac{M_{*\rm{, host}}}{M_{*\rm{, companion}}},
\end{equation}
again taking $M_*$ as the best estimate of the total stellar mass (as defined in \S{\ref{section:sample}}) of the host and companion, respectively. 

We separate our samples into three sub-categories based on stellar mass ratio: $\mu > 2$ indicates pairs with more massive hosts; $0.5 < \mu < 2$ indicates pairs with hosts and companions of similar masses; and $\mu < 0.5$ indicates pairs with more massive companions. We count the number of galaxy pairs in each mass ratio sub-category and compare the distributions of each between \textit{Hp} and \textit{Hs} pairs in both simulations. Where possible given the sample size, we compare $Q\rm(sSFR)$ for each sub-category.

%%%%%%%%%%%%%%%%%%%%%%%%%%%%%%%%%%%%%%%%%%%%%%%%%%

\section{Star Formation Enhancement by Companion Type}
\label{section:resultsMain}

\subsection{sSFR vs. 3D separation in TNG100-1}
\label{section:tng100}

\begin{figure*}
	\begin{center}
		\subfloat[TNG100-1 \textit{Hs} pairs]{
			\includegraphics[width=.45\textwidth]{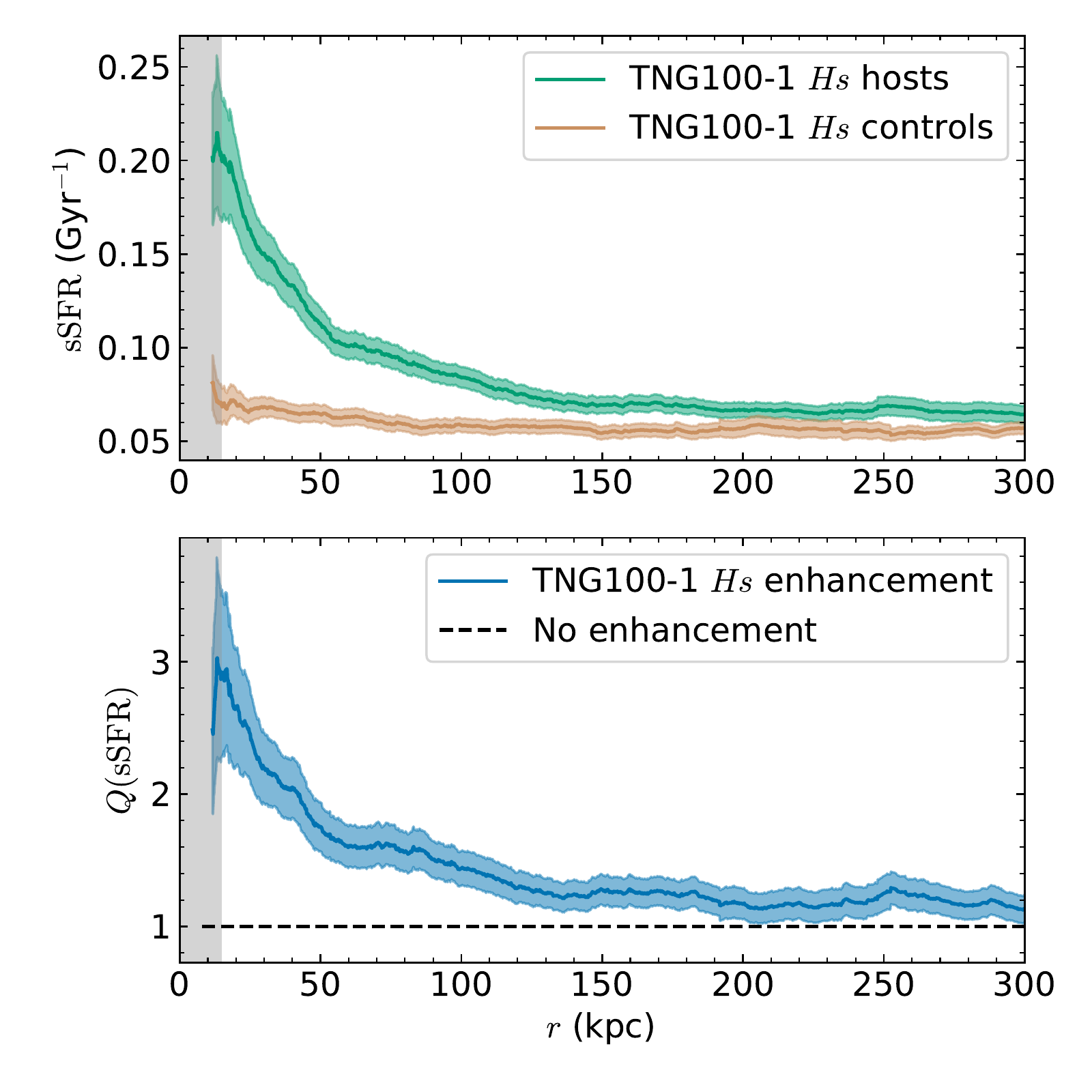}
		}
		\subfloat[TNG100-1 \textit{Hp} pairs]{
			\includegraphics[width=.45\textwidth]{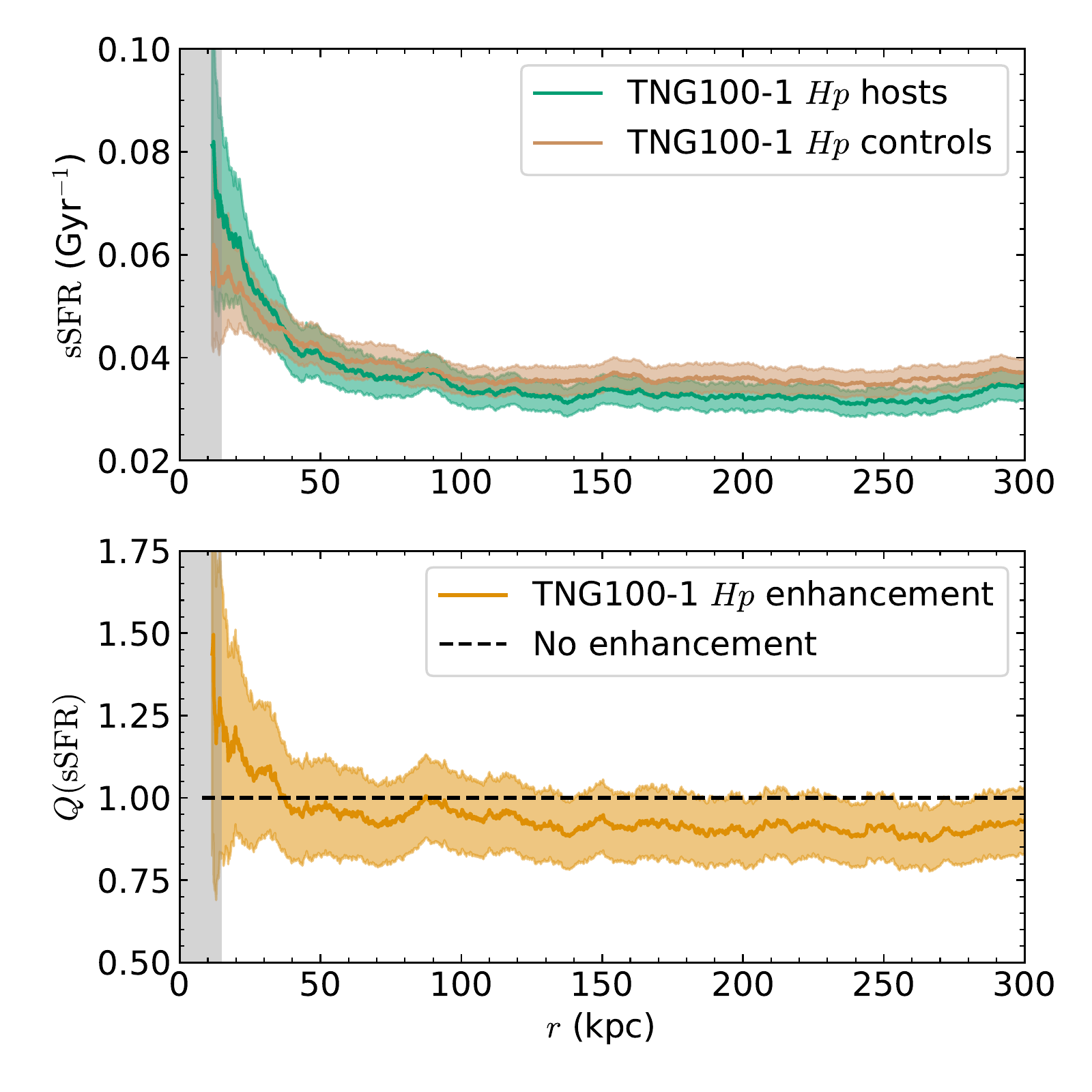}
		}
		\caption{The figures above plot the sSFR enhancement in the TNG100-1 sample as a function of galaxy pair separation for (a) \textit{Hs} pairs, and (b) \textit{Hp} pairs. In the upper panels, the mean sSFR of all host galaxies in that category (green) and their best control galaxies (tan) are plotted vs. 3D separation to the host galaxy's closest companion ($r$). In the lower panels, the sSFR enhancement of that sample ($Q$) is plotted vs. $r$, along with a black dashed line along $Q = 1$ which represents no enhancement. In all panels, the grey region represents an approximation of the crowding regime. Shaded regions in all panels represent the $2\sigma$ standard error in the mean.}
		\label{fig:t1_sSFR}
	\end{center}
\end{figure*}

\begin{figure}
	\begin{center}
		\includegraphics[width=\columnwidth]{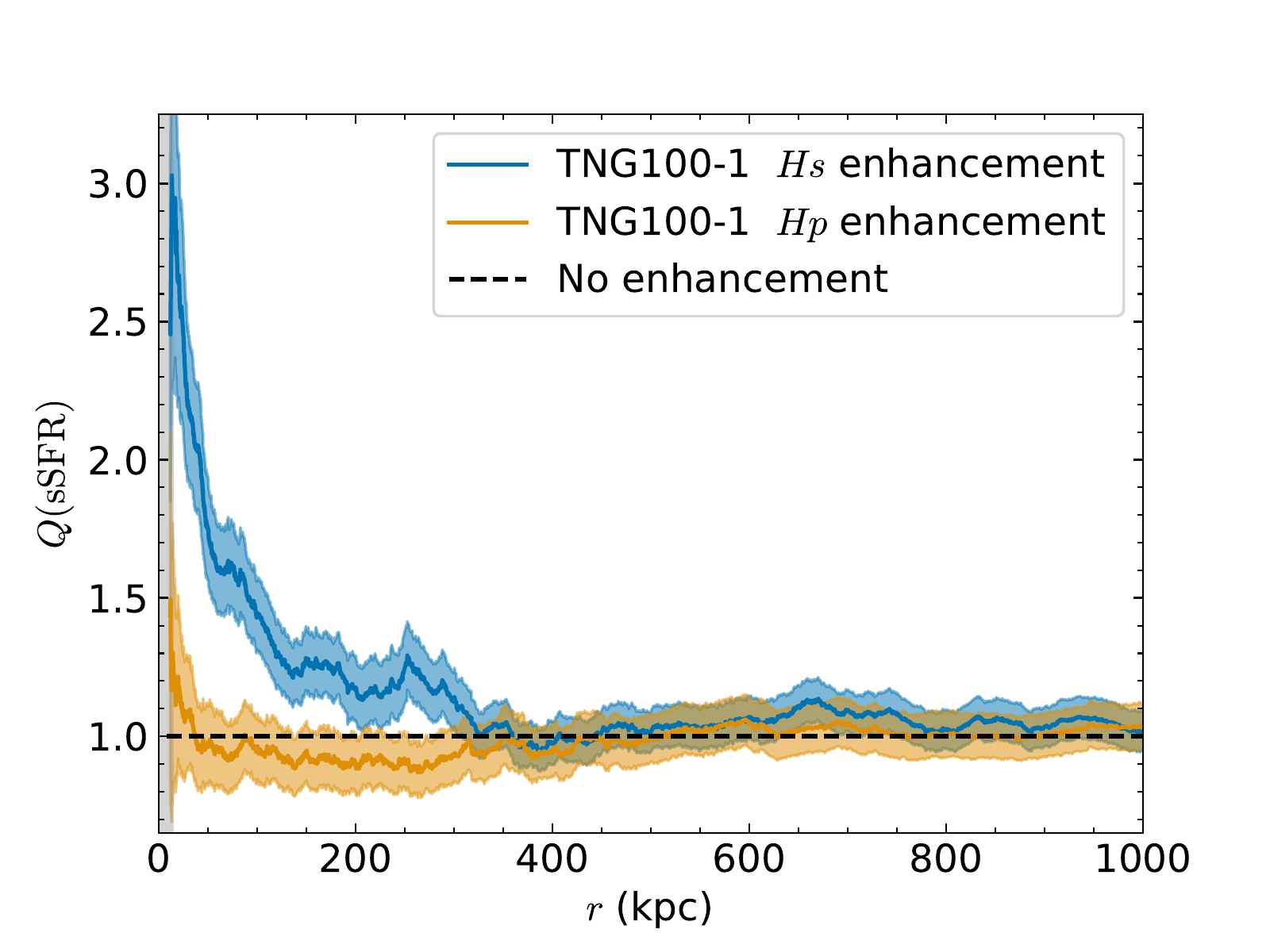}
		\caption{The sSFR enhancement ($Q$) in the TNG100-1 sample is plotted vs. 3D separation to the host's companion galaxy ($r$) for both categories of galaxy pairs, \textit{Hs} (blue), and \textit{Hp} (orange). A black dashed line is also plotted along $Q = 1$, which represents no enhancement. The grey region represents an approximation of the crowding limit. Shaded regions in both panels represent $2\sigma$ standard error in the mean.}
		\label{fig:t1_Q}
	\end{center}
\end{figure}

After dividing our original sample as described above, we have a total of 27,584 \textit{Hs} pairs and 24,190 \textit{Hp} pairs in TNG100-1 across our redshift range of $0 \leq z \leq 0.2$. We plot sSFR vs. $r$ and $Q\rm(sSFR)$ vs. $r$ out to 300 kpc in Fig. \ref{fig:t1_sSFR} for these two categories. This is supplemented by Fig. \ref{fig:t1_Q}, which plots $Q\rm(sSFR)$ for both \textit{Hs} and \textit{Hp} on the same graph for more direct comparison, out to separations of 1000 kpc.

Fig. \ref{fig:t1_sSFR}(a) shows increasing enhancements with decreasing separations for \textit{Hs} pairs, similar to what has been found for interacting galaxies in general (see Fig.~\ref{fig:West Master}). However, \textit{Hs} pairs reach a higher maximum enhancement of $Q=2.9 \pm 0.3$ at the crowding limit (compared to $Q = 2.3 \pm 0.3$), and statistically significant enhancements persist until separations as large as $r \sim 300$ kpc. In comparison to this, \textit{Hp} pairs in Fig. \ref{fig:t1_sSFR}(b) are roughly consistent with no enhancement within the $2\sigma$ confidence interval at all separations. There may be a rise in enhancement for \textit{Hp} pairs at separations of $r < 50$ kpc; however, the uncertainty at these separations is too large to confidently declare an enhancement. At the crowding limit, \textit{Hp} pairs have a maximum enhancement of only $Q = 1.2 \pm 0.2$. Both categories of galaxy pairs are consistent with no net enhancement ($Q=1$) at large separations.

The results from Fig. \ref{fig:t1_sSFR} shed new light on the overall trend of enhanced sSFR in IllustrisTNG galaxy pairs shown in Fig. \ref{fig:West Master}. It is clear that \textit{Hs} pairs are responsible for virtually all of the net sSFR enhancements that are seen, whereas \textit{Hp} pairs contribute a (small) net suppression in sSFRs.  In summary, \textit{star formation rate enhancements in galaxy pairs are largely driven by galaxies with star-forming companions.}

\subsection{sSFR vs. 3D separation in TNG300-1}
\label{section:tng300}

\begin{figure*}
	\begin{center}
		\subfloat[TNG300-1 \textit{Hs} pairs]{
			\includegraphics[width=.45\textwidth]{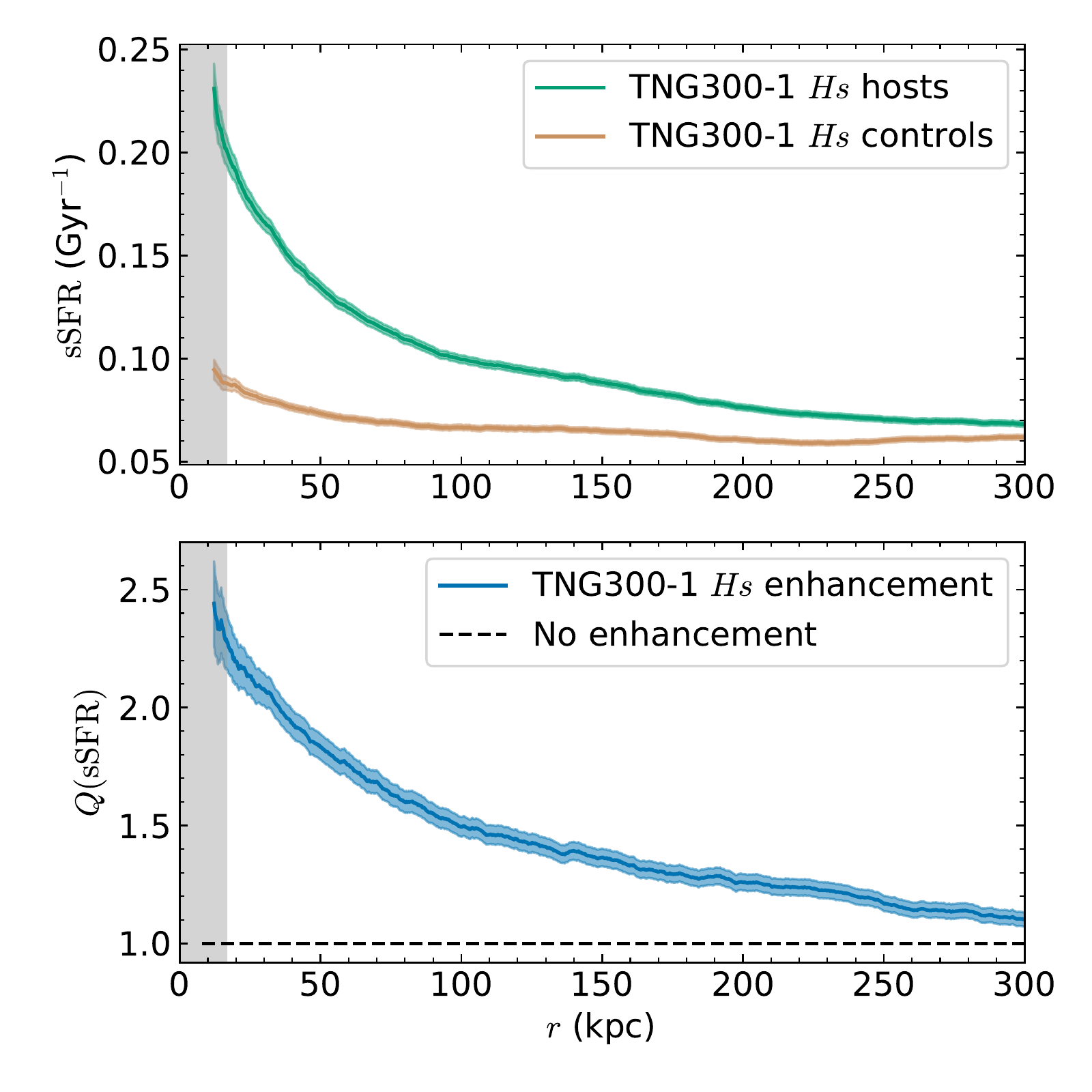}
		}
		\subfloat[TNG300-1 \textit{Hp} pairs]{
			\includegraphics[width=.45\textwidth]{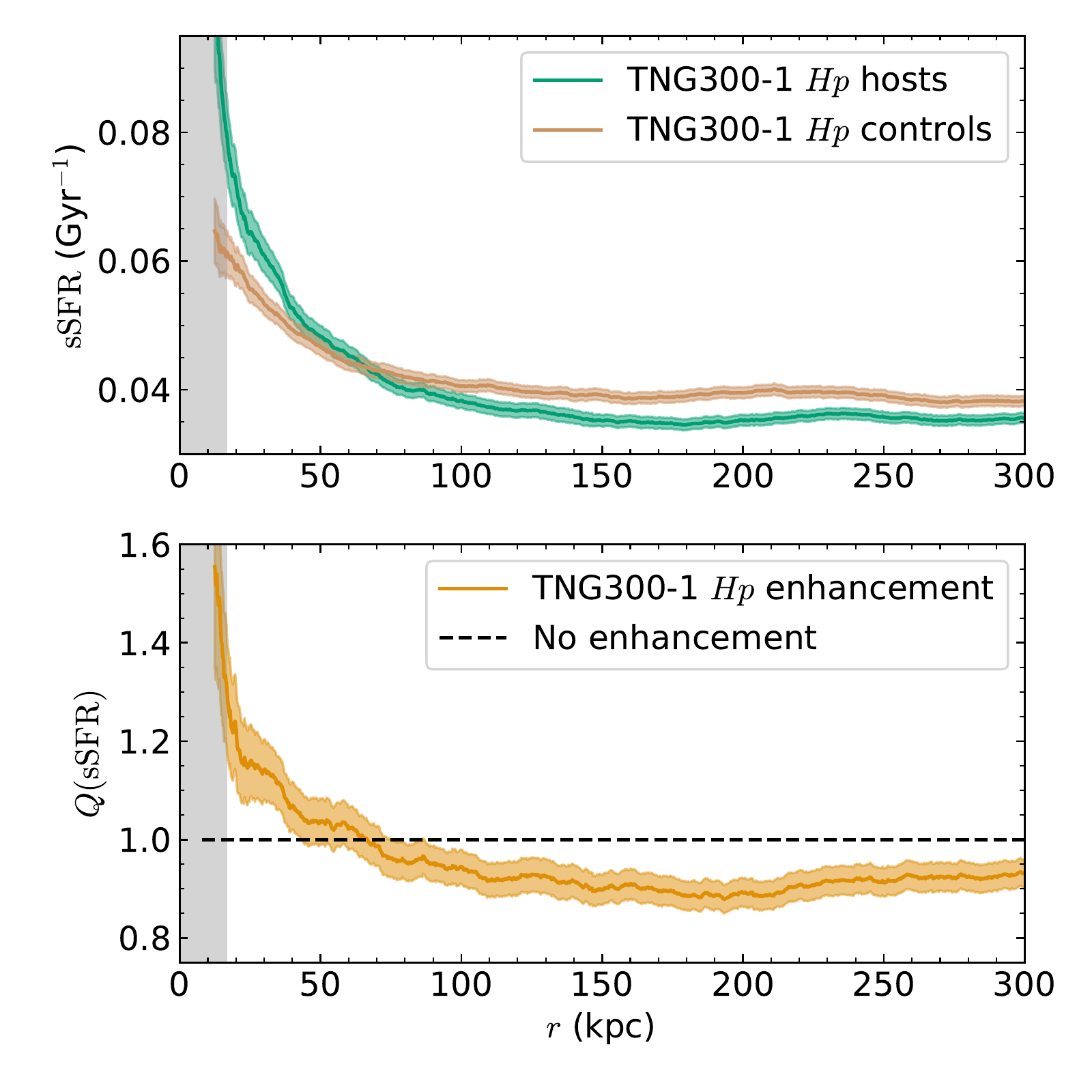}
		}
		\caption{The figures above plot the sSFR enhancement in the TNG300-1 sample as a function of galaxy pair separation for (a) \textit{Hs} pairs, and (b) \textit{Hp} pairs. In the upper panels, the mean sSFR of all host galaxies in that category (green) and their best control galaxies (tan) are plotted vs. 3D separation to the host's companion galaxy ($r$). In the lower panels, the interaction-induced sSFR enhancement of that category ($Q$) is plotted vs. $r$, along with a black dashed line along $Q = 1$ which represents no enhancement. In all panels, the grey region represents an approximation of the crowding limit. Shaded regions in all panels represent $2\sigma$ standard error in the mean.}
		\label{fig:t3_sSFR}
	\end{center}
\end{figure*}

\begin{figure}
	\begin{center}
		\includegraphics[width=\columnwidth]{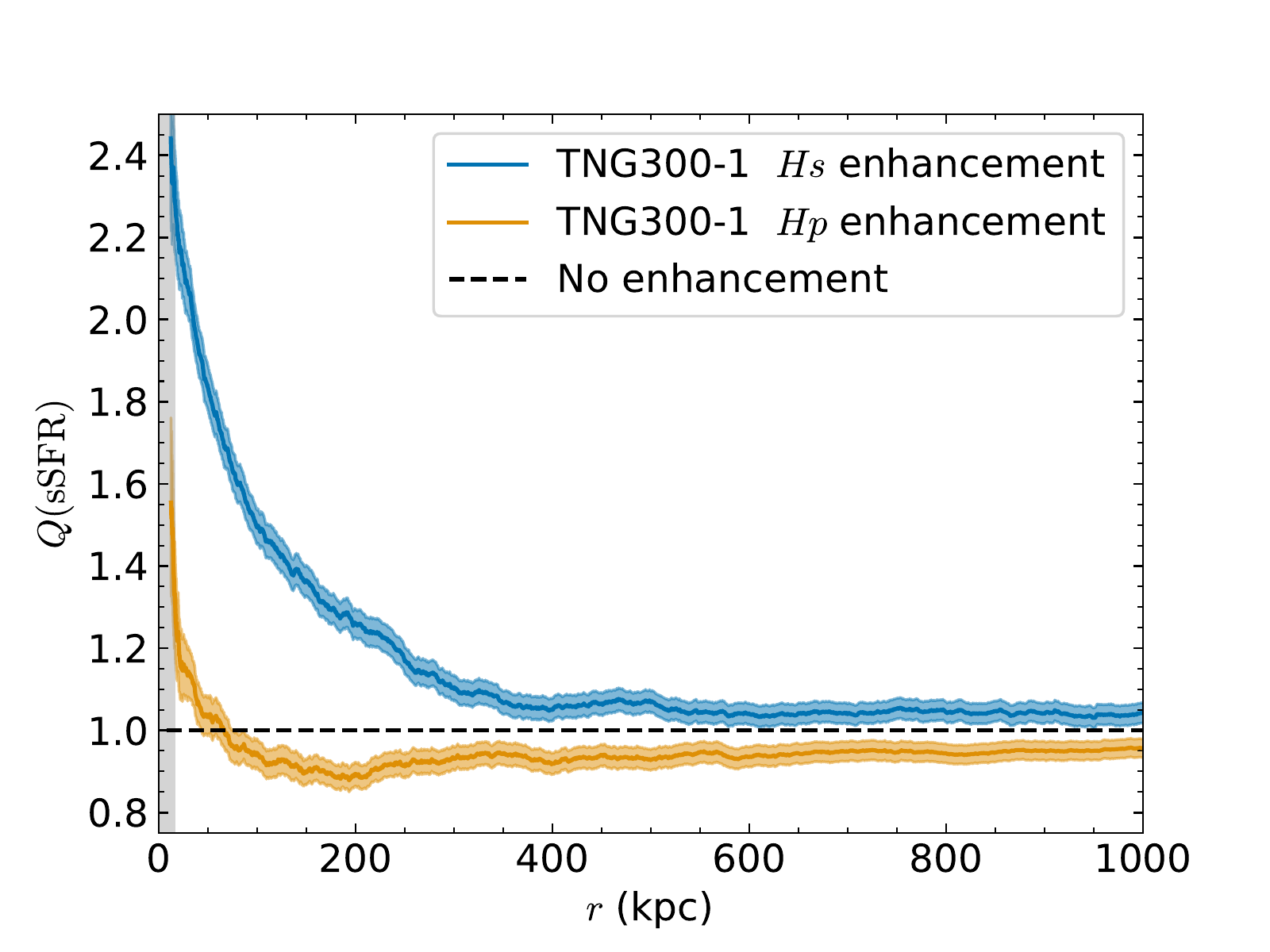}
		\caption{The sSFR enhancement ($Q$) in the TNG300-1 sample is plotted vs. 3D separation to the host's companion galaxy ($r$) for both categories of galaxy pairs: \textit{Hs} (blue) and \textit{Hp} (orange). A black dashed line is also plotted along $Q = 1$, which represents no enhancement. The grey region represents an approximation of the crowding limit. Shaded regions in both panels represent the $2\sigma$ standard error in the mean.}
		\label{fig:t3_Q}
	\end{center}
\end{figure}

Our sample of galaxies with $z \leq 0.2$ in TNG300-1 is roughly 14 times larger than our sample of galaxies in TNG100-1, resulting in 335,899 \textit{Hs} pairs and 406,092 \textit{Hp} pairs after sample separation. As with TNG100-1, we have plotted sSFR vs. $r$ of hosts and controls and $Q\rm(sSFR)$ vs. $r$ out to 300 kpc for both categories in Fig. \ref{fig:t3_sSFR}, and a combined $Q\rm(sSFR)$ vs. $r$ plot out to 1000 kpc in Fig. \ref{fig:t3_Q}.

\textit{Hs} pairs in TNG300-1, plotted in Fig. \ref{fig:t3_sSFR}(a), follow the same general trend as those in TNG100-1. Notably, TNG300-1 pairs reach lower maximum enhancements than those in TNG100-1, having only $Q=2.27 \pm 0.06$ at 17 kpc. The statistical uncertainties in the TNG300-1 sample are significantly smaller, allowing us much more certainty in the separation out to which enhancements still exist. Interaction-induced enhancements are present in \textit{Hs} pairs out to $r \sim 350$ kpc. The lower maximum enhancements and larger separations out to which enhancements exist agree with the overall results found by \citetalias{Patton2020} for TNG300-1 compared to TNG100-1, and appear to be caused by the lower resolution of the TNG300-1 simulations.

The reduced statistical uncertainties in TNG300-1 allow us to observe a more interesting trend in \textit{Hp} pairs compared with TNG100-1. At separations of $r < 50$ kpc, \textit{Hp} pairs in Fig. \ref{fig:t3_sSFR}(b) have increasingly enhanced sSFRs with decreasing galaxy separations, up to a maximum of $Q = 1.28 \pm 0.06$ at the crowding limit. At separations larger than this, notably at $100$ kpc $< r < 250$ kpc, \textit{Hp} pairs have clearly {\em suppressed} sSFRs, reaching a maximum suppression at $r \sim 200$ kpc of $Q = 0.88 \pm 0.01$. This trend may also be present in the TNG100-1 sample (see Fig.~\ref{fig:t1_Q}), but the larger statistical uncertainties in the smaller TNG100-1 simulation make it impossible to confirm this.

In Fig. \ref{fig:t3_Q}, it becomes clear that both \textit{Hs} and \textit{Hp} pairs experience a statistically significant uniform offset from $Q=1$ at large $r$, such that \textit{Hs} pairs remain slightly enhanced in sSFR and \textit{Hp} pairs remain slightly suppressed in sSFR. These offsets persist even at separations of $r \sim 1000$ kpc, and as such are unlikely to be caused by close galaxy interactions as no published studies have shown interaction effects to be observable at such large distances. It is more likely that these offsets could be evidence for galactic conformity \citep{Weinmann2006,Kauffmann2013,Knobel2015}. Galactic conformity is the phenomenon whereby galaxies in groups or clusters are more likely to share the same star-forming/passive categorization, and its effects have been shown to persist at separations as large as 4 Mpc in observations \citep{Kauffmann2013}. Within TNG300, \citet{Lacerna2022} find that an additional 24 percent of galaxies are passive when within 1 $h^{-1}$ Mpc of a passive central halo galaxy compared to the same distance from a star-forming central halo galaxy. As our control galaxies are not matched to the hosts on star-forming/passive categorization, they theoretically represent a mix of star-forming and passive galaxies. However, if our host galaxies are already shown to have a companion of one type, there may be a bias towards matching host type in cases where hosts lie in or near higher density environments. A bias towards star-forming hosts will result in net sSFR enhancement, and a bias towards passive hosts will result in net sSFR suppression when compared to an average, unbiased control group. The effect of host type is explored further in \S{\ref{section:host_type}}.

\textit{Galaxies with passive companions show signs of suppressed star formation rates.}

\subsection{Star formation enhancement dependence on mass ratio}
\label{section:mass-ratio}

Previous work has shown that the sizes of interaction-induced sSFR enhancements are also related to the difference in stellar mass between host galaxies and their companions \citep{Woods2006,Woods2007,Cox2008,Ellison2008}. To test the strength of this effect on our sample, we analyze the properties of galaxy pairs in different stellar mass ratio sub-categories (as described in \S{\ref{section:samplesep}}) and outline the results in Table \ref{tab:mratios}. In both simulations, \textit{Hs} pairs have a strong bias towards more massive host galaxies and away from more massive companion galaxies. Only 4.1 percent (3.5 percent) of \textit{Hs} pairs have $\mu < 0.5$ in TNG100-1 (TNG300-1), while 72.4 percent (70.9 percent) of \textit{Hs} pairs have $\mu > 2$. This bias is likely the result of two well-known and understood facets of galaxy populations. Firstly, there is expected to be a larger population of low mass galaxies than high mass galaxies, biasing companions towards lower masses in general. Secondly, as observational work (e.g. \citealt{Brinchmann2004}) shows, galaxies with high stellar masses ($M_* > 10^{10} M_{\odot}$) typically have lower sSFRs than moderate and low-mass galaxies, which also serves to reduce the population of star-forming companions in $\mu < 0.5$ and bias the \textit{Hs} sample towards $\mu >2$. In comparison, \textit{Hp} pairs are more evenly distributed, only slightly biased towards more massive hosts with 42.1 percent (42.5 percent) of \textit{Hp} pairs having $\mu > 2$ in TNG100-1 (TNG300-1).

Due to the smaller sample sizes in each mass ratio sub-category in TNG100-1, we choose not to proceed with further mass analysis on TNG100-1 as the statistical errors would be quite high, especially for the 1,136 \textit{Hs} pairs for which $\mu < 0.5$. All following mass analysis is performed only on the TNG300-1 samples, which still have a sufficient number of galaxy pairs from which to interpret sSFR trends over a wide range of separations.

\begin{table}
	\begin{center}
		\caption{The stellar mass ratio distribution for hosts with passive companions (\textit{Hp}) and hosts with star-forming companions (\textit{Hs}) within the TNG100-1 and TNG300-1 sample. Numbers correspond to the number of galaxy pairs within each mass ratio sub-category. Percentages correspond to the number of galaxy pairs within each mass ratio sub-category relative to the total number of galaxy pairs within that category.}
		\begin{tabular}{lc|c|c}
			\hline
			& $\mu > 2$ & $0.5 < \mu < 2$ & $\mu < 0.5$ \\
			\hline
			\textbf{TNG100-1 \textit{Hs} pairs} & 19,976 & 6,472 & 1,136 \\ 
			 & (72.4\%) & (23.5\%) & (4.1\%) \\
			\hline
			\textbf{TNG100-1 \textit{Hp} pairs} & 10,181 &  7,231 & 6,778 \\
			 & (42.1\%) & (29.9\%) & (28.0\%) \\
			\hline
			\textbf{TNG300-1 \textit{Hs} pairs} & 238,128 & 86,115 & 11,656 \\
			 & (70.9\%) & (25.6\%) & (3.5\%) \\
			\hline
			\textbf{TNG300-1 \textit{Hp} pairs} & 170,436 & 124,505 & 111,151 \\
			& (42.0\%) & (30.6\%) & (27.4\%) \\
			\hline
		\end{tabular}
		\label{tab:mratios}
	\end{center}
\end{table}

\begin{figure*}
	\begin{center}
		\subfloat[TNG300-1 \textit{Hs} pairs]{
		    \includegraphics[width=.45\textwidth]{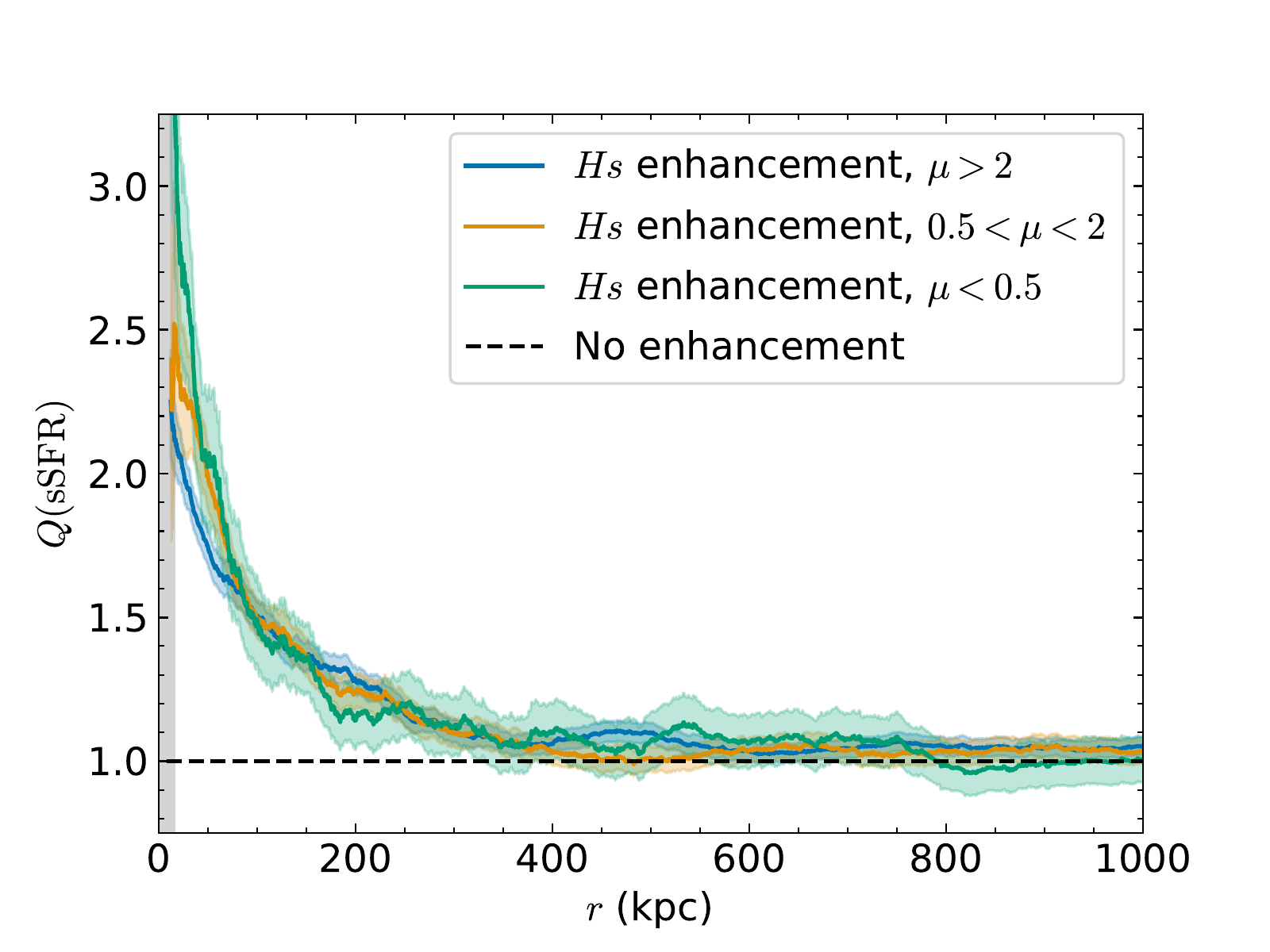}
		}
		\subfloat[TNG300-1 \textit{Hp} pairs]{
		    \includegraphics[width=.45\textwidth]{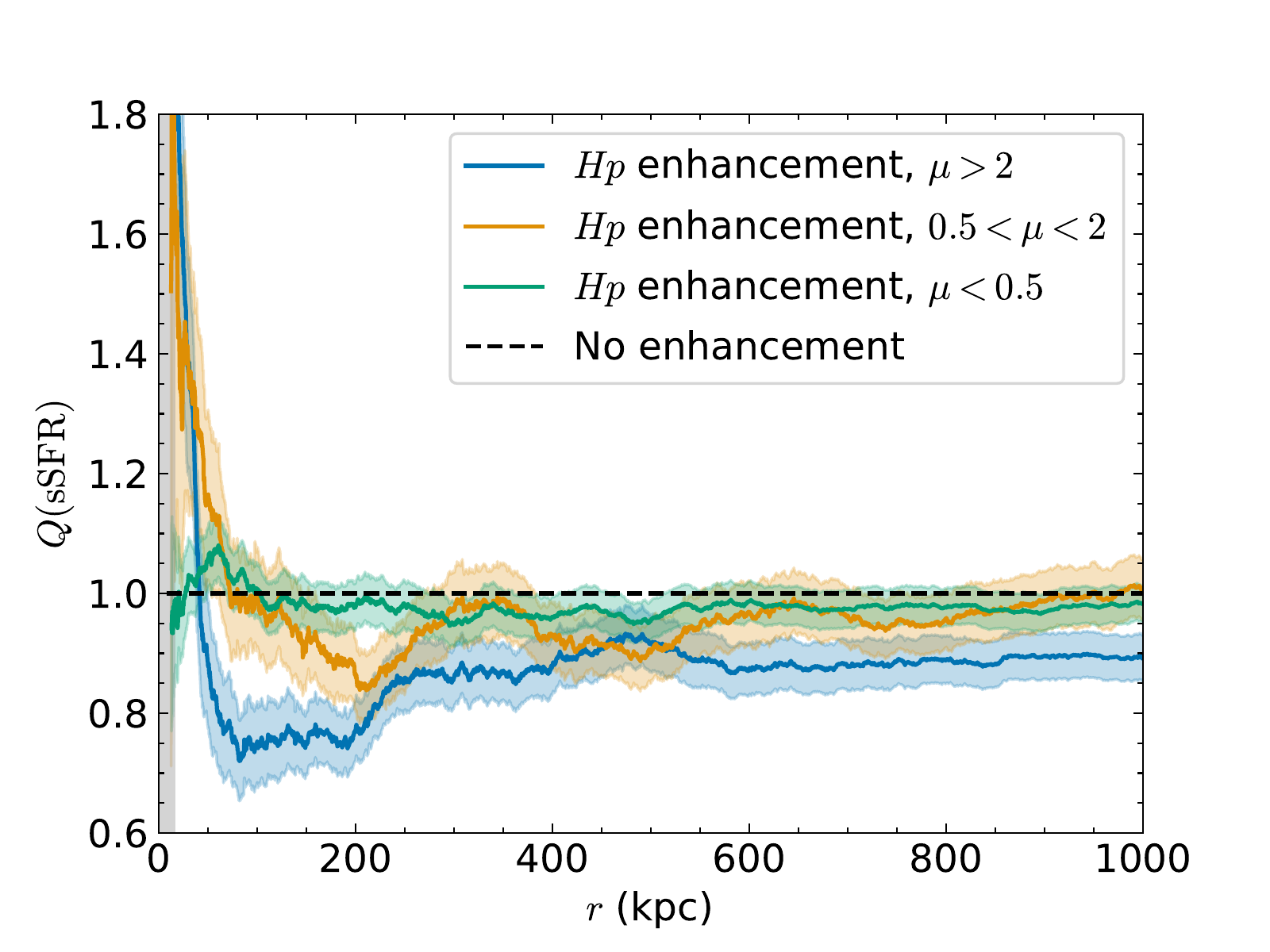}
		}
		\caption{The sSFR enhancement ($Q$) in the TNG300-1 sample is plotted vs. 3D separation to the host's companion galaxy ($r$) for (a) \textit{Hs} pairs, and (b) \textit{Hp} pairs. The samples are split into three mass ratio sub-categories: $\mu >2$ (blue), $0.5 < \mu < 2$ (orange), and $\mu < 0.5$ (green). A black dashed line is also plotted along $Q = 1$, which represents no enhancement. In both panels, the grey region represents an approximation of the crowding regime. Shaded regions represent the $2\sigma$ standard error in the mean.}
		\label{fig:mratio}
	\end{center}
\end{figure*}

In Fig. \ref{fig:mratio}(a), we have plotted $Q\rm(sSFR)$ vs. $r$ for all three mass ratio sub-categories of \textit{Hs} pairs. \textit{Hs} pairs follow nearly identical trends, with increasing enhancements at decreasing separations for all mass ratio subcategories, and only slight differences in steepness of slope. Hosts for which $\mu > 2$ reach $Q=2.12 \pm 0.06$ at the crowding limit while hosts with $0.5 < \mu < 2$ reach $Q=2.5 \pm 0.2$. Hosts for which $\mu < 0.5$, of which there are only 11,656, reach a maximum enhancement of $Q=3.2 \pm 0.3$. All sub-categories reach no enhancement by $r \sim 350$ kpc.

In comparison, \textit{Hp} pairs plotted in Fig. \ref{fig:mratio}(b) have remarkably different trends depending on the mass ratio of the host and companion galaxies. Hosts with $\mu > 2$ have increasing enhancements with decreasing separation for $r < 50$ kpc, up to $Q=2.1 \pm 0.2$ at the crowding limit. They have suppressed sSFRs between $50$ kpc $< r < 250$ kpc, with a maximum suppression of $Q=0.72 \pm 0.03$ at $r \sim 80$ kpc. Hosts for which $0.5 < \mu < 2$ have similar increasing sSFR enhancements as hosts with $\mu >2$ for $r < 75$ kpc, to a maximum enhancement of $Q=1.6 \pm 0.3$ at the crowding limit. They experience statistically significant suppressions between $150$ kpc $< r < 300$ kpc, to a minimum of $Q=0.84 \pm 0.03$ at $r \sim 200$. Unlike the other two mass ratio sub-categories, hosts with $\mu < 0.5$ are consistent with no enhancement at all galaxy separations, reaching a maximum of $Q=1.08 \pm 0.03$ near the crowding limit. These results imply that more massive hosts with smaller passive companions may be driving suppressed sSFR trends in \textit{Hp} pairs. We explore this result further in \S{\ref{section:additional_mechs}}.

%%%%%%%%%%%%%%%%%%%%%%%%%%%%%%%%%%%%%%%%%%%%%%%%%%

\section{Mass Analysis}
\label{section:mass-analysis}

The results of \S{\ref{section:mass-ratio}}, particularly for \textit{Hp} galaxies, highlight that our sSFR enhancements are significantly mass-dependent. Therefore, if we wish to specifically analyze the influence of star-forming vs passive companions, and not simply companions of different masses, we must seek to eliminate mass differences between our samples of \textit{Hs} and \textit{Hp} pairs.

\subsection{Creating a "twinned" mass sample}
\label{section:methods-twinned}

In our TNG300-1 samples, where our sample size is quite large, we take our mass analysis a step further by removing all stellar mass differences between our \textit{Hp} and \textit{Hs} pairs, effectively creating a "twinned" sample. To do this, we design a process whereby we begin with the sample of \textit{Hs} pairs and attempt to find a pairwise stellar mass match for each one among the \textit{Hp} pairs. The choice of beginning with the \textit{Hs} pairs instead of the \textit{Hp} pairs is based on the distribution of our pair sample, and ultimately reduces the number of pairs with no matches or repeated matches. The choice of which category of pairs to begin with does not have a significant impact on our results.

Galaxy pairs are matched on host stellar mass, companion stellar mass, and $r$. We do not attempt to match galaxy pairs on environment, as we note that our control galaxies are already matched on two environmental factors ($N_2$ and $r_2$). We require all stellar masses to agree within 0.05 dex and $r$ to agree within 10 percent. We do not require a match on snapshot, as our snapshot range is already quite limited as to significantly reduce the differences between epochs. We use weighting terms to determine the best simultaneous match on all parameters, as described in \citet{Patton2016}.

As we choose to only keep a single galaxy pair as the best match, we introduce an additional weighting term to reduce the frequency of repeated matches (when one \textit{Hp} pair is chosen as the best match for multiple \textit{Hs} pairs).  This term guards against biasing the matched sample towards a handful of specific galaxy pairs. We call this the repeat tolerance term, and it is defined as follows:
\begin{equation}
    w\textsubscript{repeat}_i = 0.9^{n_i},
\end{equation}
where $n_i$ is the number of times the  $i^{\rm th}$ galaxy pair has already been chosen as a best match. The factor of 0.9 can be varied depending on the preferred tolerance for repeated matches, but we find that varying this factor does not significantly alter our results. Controlling the repeat tolerance allows us to maintain the largest possible sample size compared to eliminating repeated matches, while encouraging a more diverse sample of high-quality matches. The pairs with the highest number of repeated matches after including the repeat tolerance term are predominantly at large separations ($r > 800$ kpc) where the sample size is smaller but the range of possible matched $r$ values is larger.

If a match is not found for an \textit{Hs} pair within the initial parameter limits, it is removed from the twinned sample. Repeated \textit{Hp} pairs are treated as multiple independent galaxy pairs to ensure that the number of twinned \textit{Hs} and \textit{Hp} pairs remains equal. In order to ensure that all matches are plotted in the same bins for averaging, all \textit{Hp} pairs are plotted at the $r$ value of the \textit{Hs} pair they are matched to.

\subsection{Results from the twinned sample}
\label{section:results-twinned}

\begin{figure}
	\begin{center}
		\includegraphics[width=\columnwidth]{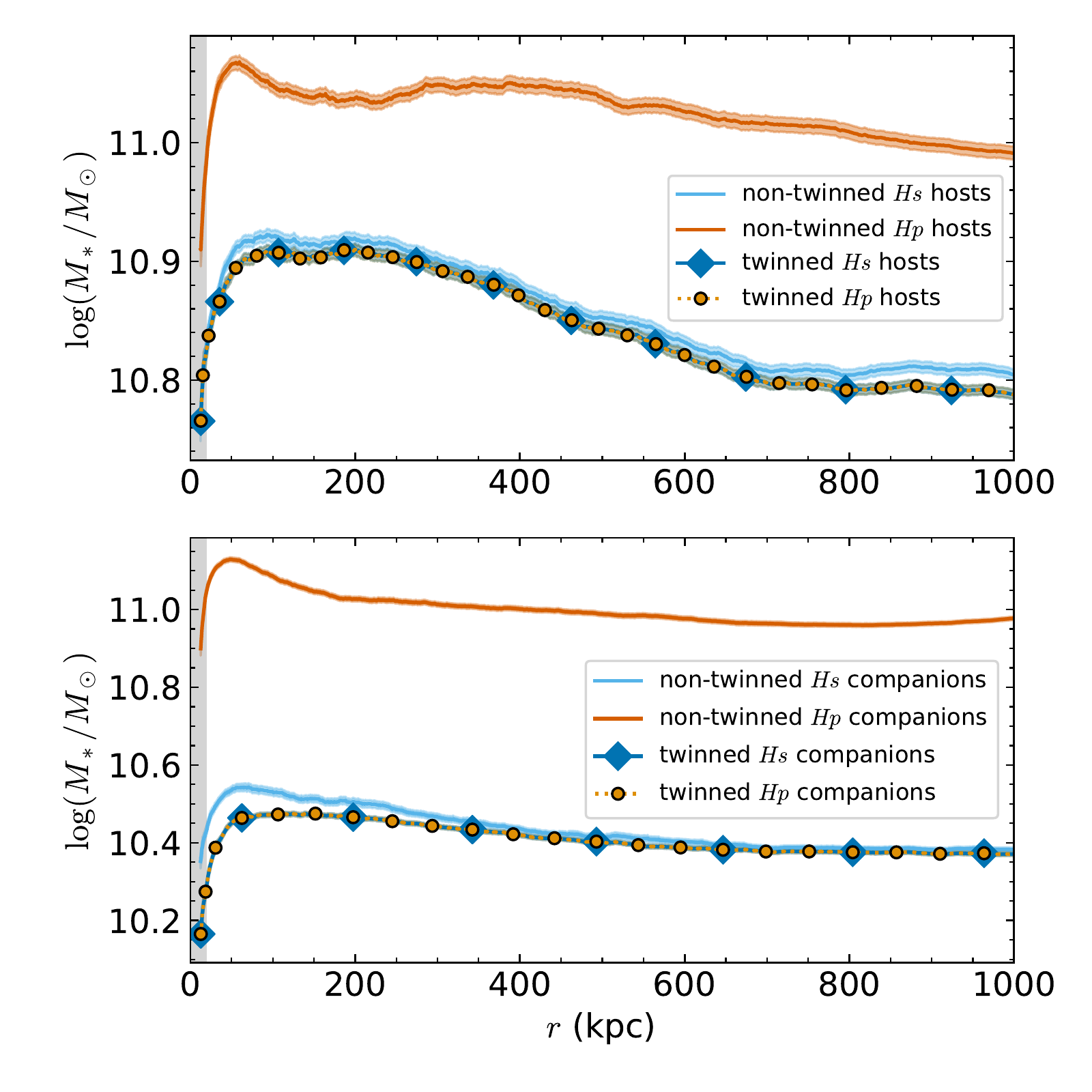}
		\caption{The total stellar mass of the host galaxy (top panel) and the total stellar mass of the companion galaxy (bottom panel) are plotted vs. the 3D separation to the host's companion galaxy ($r$). Masses of the initial, non-twinned TNG300-1 \textit{Hs} (light blue) and \textit{Hp} (dark orange) galaxies are plotted as solid lines. Masses of twinned \textit{Hs} pairs (blue) are plotted with diamond markers and their best \textit{Hp} match (orange) are plotted with circle markers to distinguish the overlapping trends. The grey region represents an approximation of the crowding limit. Shaded regions in both panels represent $2\sigma$ standard error in the mean.}
		\label{fig:mass comp}
	\end{center}
\end{figure}

To ensure that the mass-matching of our twinned sample is reliable, we plot the total stellar mass vs. $r$ for all initial non-twinned and twinned \textit{Hs} and \textit{Hp} galaxies. Masses of host galaxies are plotted in the upper panel of Fig. \ref{fig:mass comp}, while masses of companion galaxies are plotted in the lower panel. The twinned \textit{Hs} galaxy masses (blue diamonds) overlap almost perfectly with the \textit{Hp} galaxies (orange circles) at all separations, indicating that our matching process has returned twinned samples that have virtually indistinguishable stellar mass distributions. We note, as in \citetalias{Patton2020}, that the stellar masses of our hosts and companions do show a dependence on $r$. The decrease in stellar mass at the smallest separations is an expected outcome of \citetalias{Patton2020}'s removal of overlapping galaxies from their pair sample---massive galaxies will have more extended $R_{1/2}$ and therefore overlap at separations larger than those of less massive galaxies. We note that the differences in stellar mass between close separations and wide separations are only $\sim$ 0.1 dex for both host galaxies and companion galaxies, and are therefore not considered significant.

Our matching process yielded successful matches for 97.5 percent of our original TNG300-1 \textit{Hs} sample. This brings our new twinned sample to 327,573 \textit{Hs} pairs and 327,573 \textit{Hp} pairs with identical mass distributions, leaving the star-forming/passive categorization of the companions as the main difference between the samples . The matching process fails to find a match more frequently at small separations, leading to more unreliable averages at small separations due to small sample sizes, so we extend the crowding limit to $r=20$ kpc for our twinned samples. We note that the twinned \textit{Hp} sample consists of 166,904 unique pairs, and each pair is chosen as the best match an average of 1.96 times (median of 2).

\begin{figure}
	\begin{center}
		\includegraphics[width=\columnwidth]{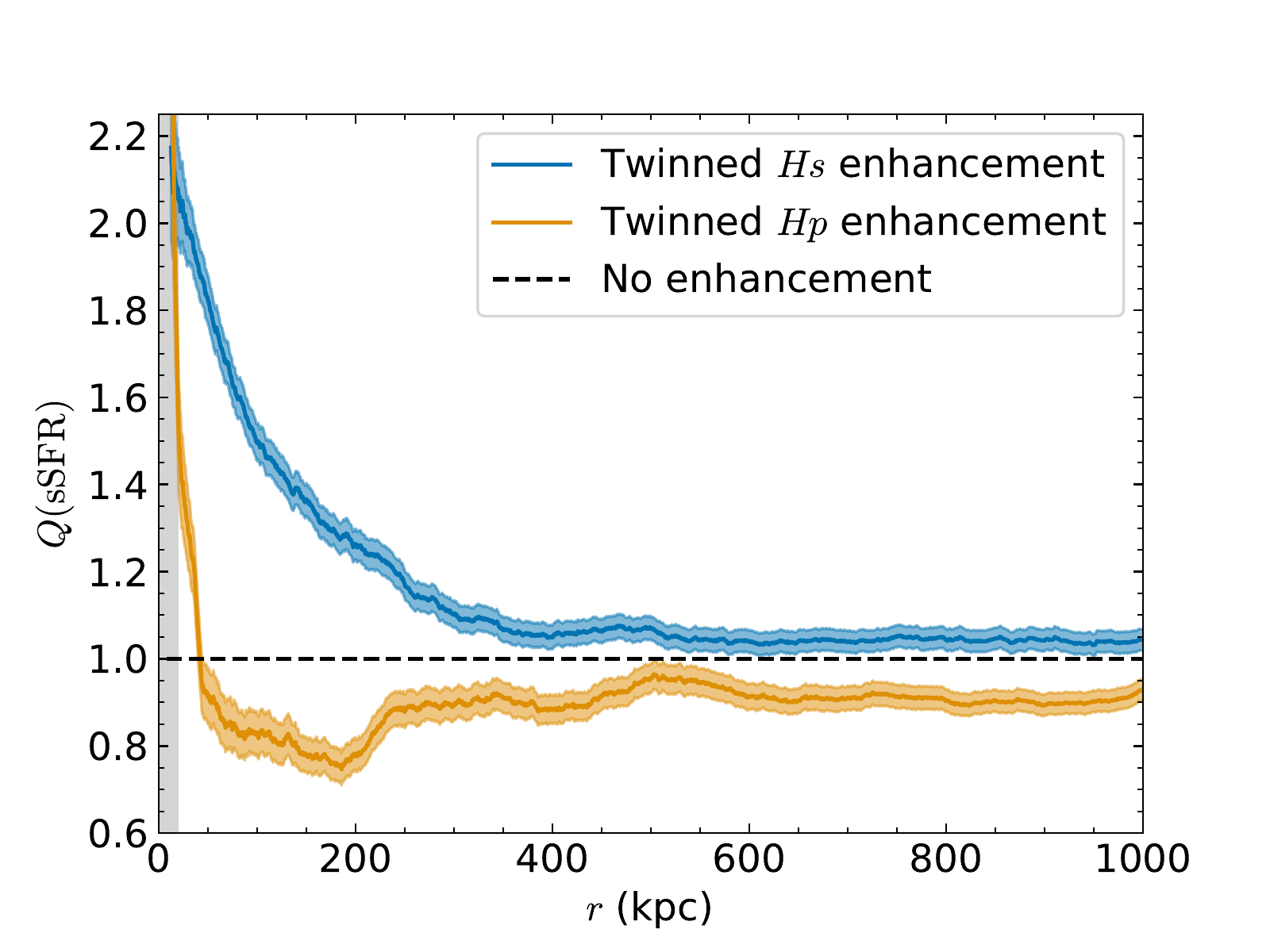}
		\caption{The sSFR enhancement ($Q$) in the twinned TNG300-1 sample is plotted vs. 3D separation to the host's companion galaxy ($r$) for both categories of galaxy pairs, \textit{Hs} pairs (blue), and \textit{Hp} pairs (orange). A black dashed line is also plotted along $Q = 1$, which represents no enhancement. The grey region represents an approximation of the crowding limit. Shaded regions in both panels represent $2\sigma$ standard error in the mean. See Fig. \ref{fig:t3_Q} for the equivalent untwinned sample.}
		\label{fig:mass-match_Q}
	\end{center}
\end{figure}

In Fig. \ref{fig:mass-match_Q}, we plot $Q\rm(sSFR)$ vs. $r$ for our twinned \textit{Hs} pairs and \textit{Hp} pairs. The twinned \textit{Hs} pairs follow the familiar trend of increasing $Q$ with decreasing $r$, reaching a maximum enhancement of $Q=2.06 \pm 0.05$ at the new crowding limit of $r=20$ kpc, with enhancements present out to separations as large as $r \sim 350$ kpc. Twinned \textit{Hp} pairs have increasingly enhanced sSFRs at small separations where $r < 50$ kpc, up to a maximum of $Q=1.52 \pm 0.07$ near the crowding limit. Between $50$ kpc $< r < 250$ kpc, twinned \textit{Hp} pairs are suppressed with enhancements as low as $Q=0.75 \pm 0.02$ (around $r \sim 185$ kpc). At large separations, from $r > 600$ kpc, both \textit{Hs} and \textit{Hp} galaxies have uniform offsets such that \textit{Hs} pairs are slightly enhanced and \textit{Hp} pairs are slightly suppressed, as with the non-twinned TNG300-1 samples. We again believe that this is likely to be an indication of galactic conformity, as discussed in \S{\ref{section:tng300}}. All of these trends are remarkably similar to those in our untwinned sample (Fig.~\ref{fig:t3_Q}), demonstrating conclusively that these trends are not driven by differences in the stellar mass distributions of \textit{Hs} and \textit{Hp} pairs.

We note that the twinned sample is a useful diagnostic tool to rule out intrinsic stellar mass differences as the primary driver behind our sSFR differences, but that it would be inadvisable to perform additional analysis or derive further results. For example, although running an analysis of sSFR enhancement for different mass ratios within the twinned \textit{Hp} sample does result in different trends to those found in Fig.~\ref{fig:mratio}(b), we elect not to include this plot in the paper as it cannot be used to draw physical conclusions on the general \textit{Hp} population. As seen in Fig.~\ref{fig:mass comp}, the twinned sample excludes many of the high mass pairs contained in the full \textit{Hp} sample. The twinned \textit{Hp} sample is therefore only representative of a sub-sample of galaxy pairs which are most similar in mass distribution to \textit{Hs} pairs.

Using the twinned sample as a diagnostic check, we are able to conclude that \textit{intrinsic galaxy mass differences are not responsible for the differences in star formation rates between galaxies with star-forming companions and galaxies with passive companions.}

%%%%%%%%%%%%%%%%%%%%%%%%%%%%%%%%%%%%%%%%%%%%%%%%%%

\section{Connecting Simulations to Observations}
\label{section:observational}

We stated in the introduction that cosmological simulations are useful for the study of galaxies so long as they are a decent representation of the real physics of the universe. Therefore, in order to draw any useful physical conclusions from simulated data, we must check that the results of our study are reproducible in the observed universe. To that end, we proceed by creating comparable samples using the IllustrisTNG cosmological simulation and the Sloan Digital Sky Survey (SDSS).

\subsection{Projection effects}
\label{section:projection-effects}

One of the strengths of using cosmological simulations is our ability to know precise 3D locations, and by extension to know the true  separation between galaxies. In real observations, however, we are limited to observing galaxies from our viewpoint on (or near) Earth. The inherent differences in conducting observational studies means that we may receive different results depending on how sensitive our data is to projection effects. As such, we test the sensitivity to projection effects in IllustrisTNG galaxies by repeating our study using artificially projected galaxy separations ($r_{\rm p}$) instead of 3D separations.

We use the same projected samples as in \citetalias{Patton2020}. 3D galaxy locations are first projected onto the $x-y$ plane. New companions and controls are found for all hosts using projected separation rather than 3D separation.  We impose a maximum relative velocity along the line of sight of $\Delta v < 1000$ km s$^{-1}$ for all potential companions. Afterwards, all pairs for which $\Delta v > 300$ km s$^{-1}$ are eliminated from the sample. The process of creating this sample using projected separations is described in more detail in \citetalias{Patton2020}.

\begin{figure}
	\begin{center}
		\includegraphics[width=\columnwidth]{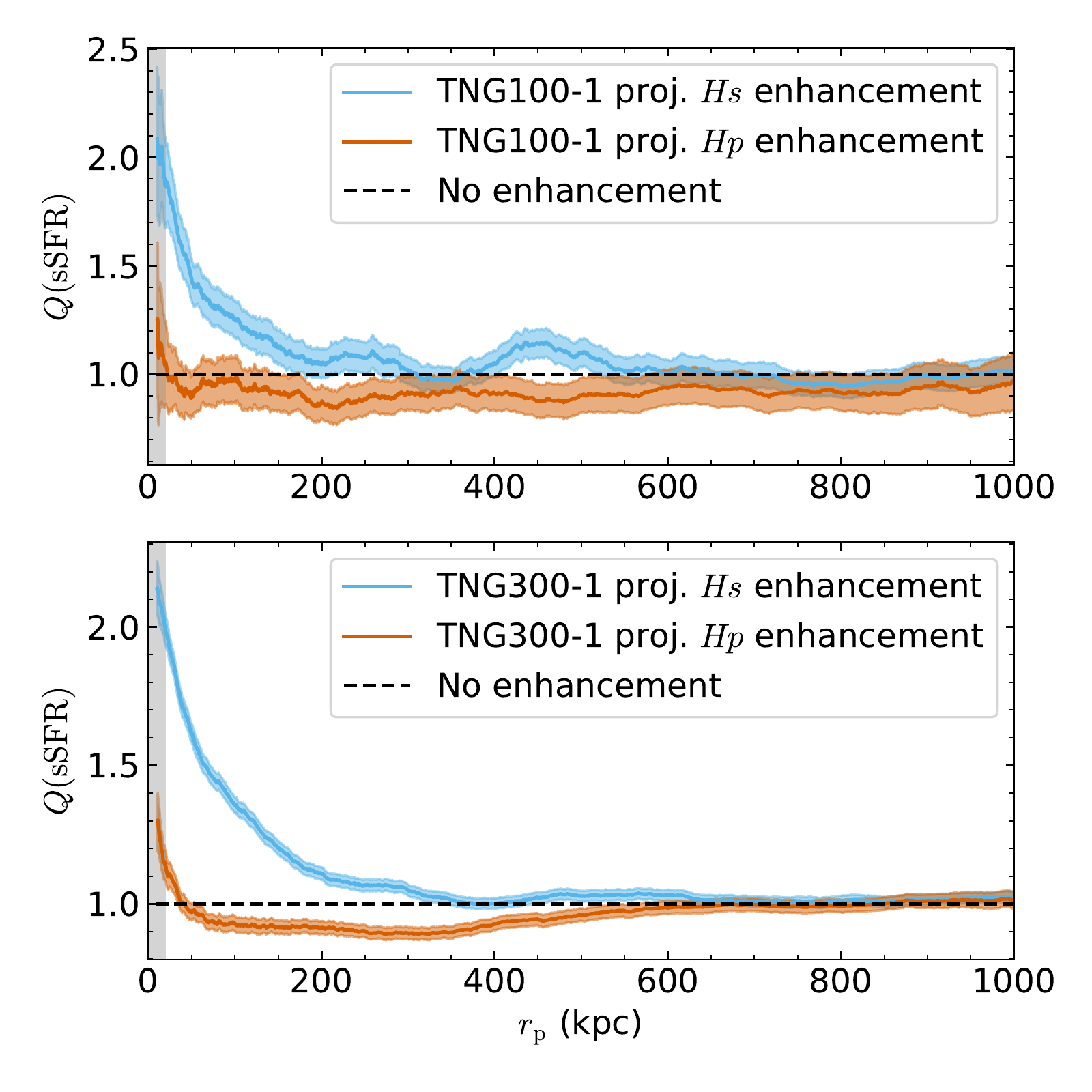}
		\caption{The sSFR enhancement ($Q$) in the TNG100-1 (top panel) and TNG300-1 (bottom panel) samples are plotted vs. projected separation to the host's companion galaxy ($r_{\rm p}$) for both categories of galaxy pairs, \textit{Hs} pairs (light blue), and \textit{Hp} pairs (dark orange). A black dashed line is also plotted along $Q = 1$, which represents no enhancement. The grey region represents an approximation of the crowding limit. Shaded regions in both panels represent $2\sigma$ standard error in the mean. See Fig. \ref{fig:t1_Q} and Fig. \ref{fig:t3_Q} for the equivalent samples using 3D separation.}
		\label{fig:projected}
	\end{center}
\end{figure}

After restricting our sample to $z \leq 0.2$ and separating our sample by companion type, we are left with 33,336 \textit{Hs} pairs and 18,657 \textit{Hp} pairs in TNG100-1, as well as 436,032 \textit{Hs} pairs and 325,535 \textit{Hp} pairs in TNG300-1. We note that there is a higher proportion of projected \textit{Hs} pairs than \textit{Hp} pairs compared to the same 3D categories. The combined $Q\rm(sSFR)$ vs. $r_{\rm p}$ plots for both simulations are shown in Fig. \ref{fig:projected}.

The results of TNG100-1 and TNG300-1 using projected separation do not vary significantly from the results using 3D separation. TNG100-1 \textit{Hs} pairs reach a maximum enhancement of $Q=2.0 \pm 0.1$ at the crowding limit, decreasing to no enhancement at projected separations of $r_{\rm p} \sim 300$ kpc, while \textit{Hp} pairs are consistent with no enhancement. While we do note a small bump in enhancement of TNG100-1 \textit{Hs} pairs at $r \sim 450$ kpc, this excess is not present in TNG300-1. Since TNG300-1 yields better control matching than TNG100-1 (since there are many more controls available per galaxy in the larger volume), we attribute this small excess to imperfect control matching in TNG100-1, rather than a  physically significant effect. TNG300-1 \textit{Hs} pairs have a similar maximum enhancement of $Q=2.00 \pm 0.03$ at the crowding limit, decreasing to no enhancement at $r_{\rm p} \sim 350$ kpc. TNG300-1 \textit{Hp} pairs have small enhancements for $r_{\rm p} < 50$ kpc, to a maximum of $Q=1.14 \pm 0.03$ at the crowding limit, but have suppressed sSFRs between $50$ kpc $< r_{\rm p} < 450$ kpc, with a maximum suppression of $Q=0.89 \pm 0.01$ at $r_{\rm p} \sim 325$ kpc.

Lastly, both TNG100-1 and TNG300-1 projected samples appear to reach no enhancement by the largest separations (except for the TNG100-1 Hp pairs, which are slightly lower than $Q=1$ and only barely consistent with no enhancement within $2\sigma$). Uniform sSFR offsets at large separations, such as $r \sim 1000$ kpc, are not present in the TNG300-1 projected sample as they are in the 3D sample. We believe this to be the result of contamination from non-physical pairs due to the selection of companions in projected space (i.e. a host whose closest 3D companion may be star-forming, but closest projected companion is passive), which would dilute the observed strength of galactic conformity.

\subsection{Comparison between TNG300-1 and SDSS}
\label{section:sdss}

As with \citetalias{Patton2020}, we use our projected sample to facilitate a direct comparison between simulated galaxy samples from IllustrisTNG and observed galaxy samples from SDSS. We begin with the sample of SDSS galaxy pairs used by \citet{Patton2013}. This sample of galaxy pairs comes from the SDSS Data Release 7 \citep{Abazajian2009}. We use the total stellar mass measurements of \citet{Mendel2014}, which are derived using galaxy photometry from \citet{Simard2011}. We use star formation rate estimates from \citet{Brinchmann2004}.

We restrict our redshift range to $0.02<z<0.2$, which is similar to our redshift cuts in IllustrisTNG. We impose a minimum stellar mass of $10^{10} M_{\odot}$, although we note that the original SDSS sample is flux-limited, which may still result in incompleteness. Specifically, our companion sample may be incomplete down to 10 percent of the host mass. sSFR is calculated using fibre SFRs from \citet{Brinchmann2004} and fibre stellar masses by \citet{Mendel2014}. SDSS fibres have a typical covering fraction of 30 percent, which varies  depending on redshift and galaxy size \citep{Patton2011}. Following \citetalias{Patton2020}, we consider SDSS sSFR calculated within fibres to be approximately analogous to sSFRs within $R_{1/2}$ in IllustrisTNG. Unlike \citetalias{Patton2020}, we do not require a minimum sSFR for our pair sample.

Closest companions and control galaxies are found using the methodology from \citet{Patton2016}, where a weighting scheme is used for incompleteness, fibre collisions, and redshift dependent mass completeness \citepalias{Patton2020}. Unlike our IllustrisTNG sample, the SDSS hosts are compared to at least 10 control galaxies, using weighting terms for best matches on redshift, stellar mass, local density, and isolation. We note that due to the limited galaxy samples available from simulations, it becomes extremely difficult to find multiple well-matched controls for our IllustrisTNG galaxies, hence this method of control matching is only viable for the observational SDSS sample. As in our projected IllustrisTNG samples, we remove galaxy pairs for which $\Delta v > 300$ km s$^{-1}$. We direct the reader to \citetalias{Patton2020} and \citet{Patton2016} for more details regarding SDSS sample generation up to this point.

\begin{figure}
	\begin{center}
		\includegraphics[width=\columnwidth]{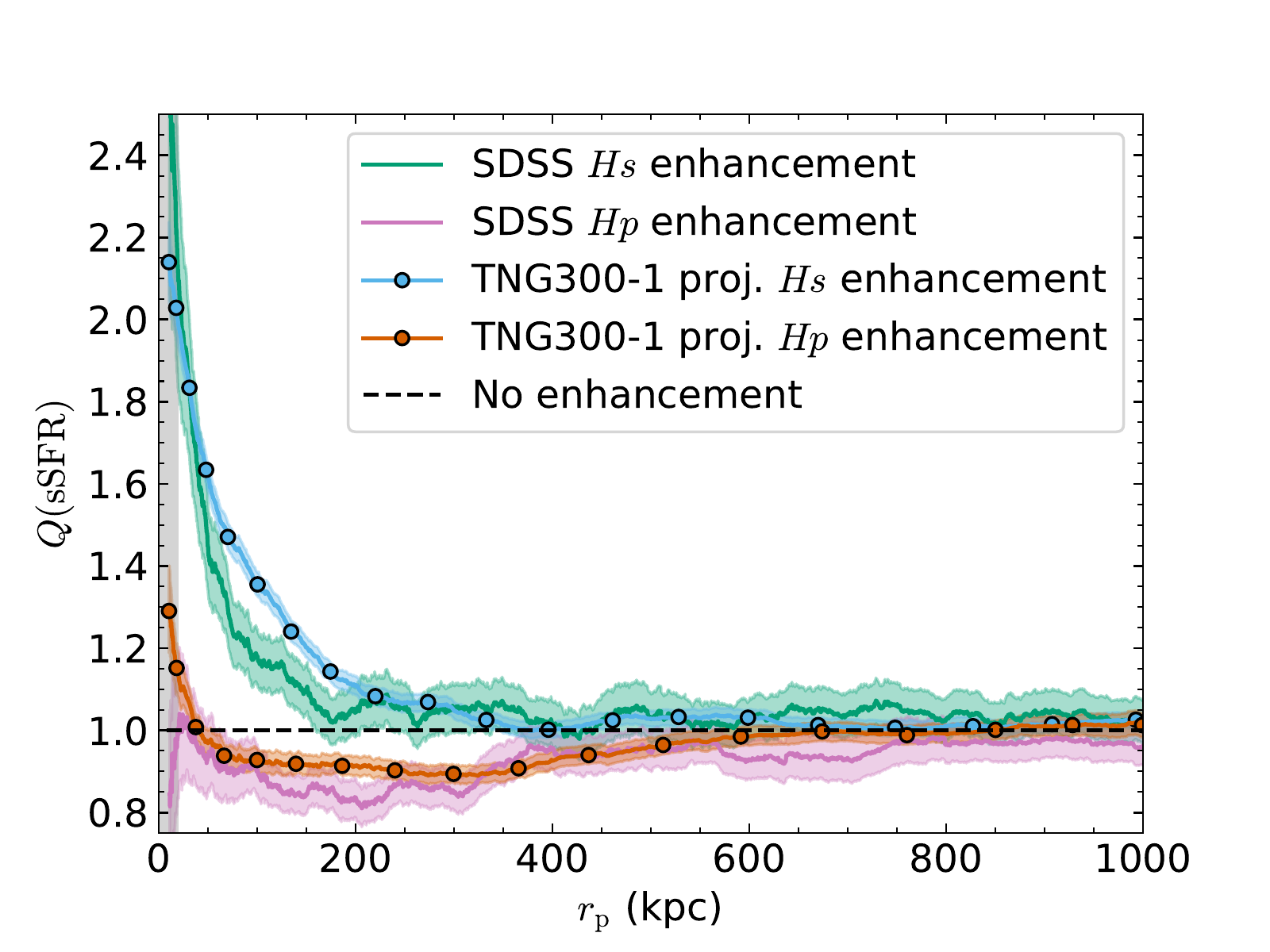}
		\caption{The sSFR enhancement ($Q$) in the SDSS \textit{Hs} (green) and \textit{Hp} (pink) samples and the projected TNG300-1 \textit{Hs} (light blue with circles) and \textit{Hp} (dark orange with circles) samples are plotted vs. projected separation to the host's companion galaxy ($r_{\rm p}$). A black dashed line is also plotted along $Q = 1$, which represents no enhancement. The grey region represents an approximation of the crowding limit. Shaded regions in both panels represent $2\sigma$ standard error in the mean.}
		\label{fig:sdss}
	\end{center}
\end{figure}

Finally, we separate our SDSS sample into \textit{Hs} and \textit{Hp} galaxies using a threshold of 0.01 Gyr$^{-1}$ sSFR. This results in 93,929 \textit{Hs} pairs and 108,987 \textit{Hp} pairs from SDSS. We directly compare the sSFR enhancement between SDSS samples and TNG300-1 samples in Fig. \ref{fig:sdss} and find generally good agreement between the two samples. We note that a similar plot for TNG100-1 (not shown) also shows good agreement with the SDSS sample, but the much larger uncertainties in TNG100-1 (due to the smaller sample size) make it impossible to identify meaningful differences between the samples. The sample of SDSS \textit{Hs} galaxies reaches a maximum enhancement of $Q=2.1 \pm 0.1$ at $r_p = 20$ kpc, while the SDSS \textit{Hp} galaxies are consistent with no enhancement or suppression with $Q=0.99 \pm 0.02$ at $r_p=20$ kpc, and reach maximum sSFR suppressions of $Q=0.81 \pm 0.02$ at $r_p \sim 200$ kpc.

While the trends of both samples are similar, we point out that the projected TNG300-1 sample (indicated by the lines with circular markers) does differ slightly from the SDSS sample. Between $50$ kpc $< r_{\rm p} < 200$ kpc, \textit{Hs} galaxies have somewhat higher sSFR enhancements than those of equivalent SDSS galaxies--or, equivalently, TNG300-1 \textit{Hs} pairs have a more gradual slope in sSFR enhancement compared to SDSS \textit{Hs} pairs. We also note that while our TNG300-1 \textit{Hp} galaxies have sSFRs that are statistically enhanced (by a small amount) for $r_{\rm p} < 50$ kpc, equivalent SDSS \textit{Hp} galaxies return to non-suppressed sSFR levels without becoming noticeably enhanced at the smallest separations. TNG300-1 \textit{Hp} galaxies may also have sSFRs that are slightly less suppressed than their counterparts in SDSS, as well as exhibiting a more gradual shift between suppressed and neutral sSFR levels between $400$ kpc $< r_{\rm p} < 600$ kpc compared to the steeper shift from $300$ kpc $< r_{\rm p} < 375$ kpc in SDSS.

\textit{There is good agreement between star formation rate trends of galaxy pairs in IllustrisTNG and SDSS.}

%%%%%%%%%%%%%%%%%%%%%%%%%%%%%%%%%%%%%%%%%%%%%%%%%%

\section{Star Formation Enhancement by Host Type}
\label{section:host_type}

\subsection{Separation of sample by host type}
\label{section:host_type_methods}

While the focus of this paper is largely on the effect of companion type on the star formation rates of interacting galaxies, it is worthwhile to take our analysis a step further and examine the possible impacts of host type, as done by \citet{Moon2019}. To that end, we take our larger original 3D TNG300-1 samples (from \S{\ref{section:tng300}}), which have already been separated by companion type, and add an additional layer of separation by host type. We use the same sSFR threshold of 0.01 Gyr$^{-1}$ to delineate between star-forming and passive host galaxies, as we did with the companion galaxies. With this, we construct samples of only star-forming hosts by reducing our 3D TNG300-1 \textit{Hs} and \textit{Hp} samples to 181,296 \textit{Ss} and 122,209 \textit{Sp} pairs. We avoid direct analysis of passive hosts due to difficulties measuring sSFR of passive galaxies, but note that we return to exploring them in \S{\ref{section:passive_frac}}. We repeat this process with our projected TNG300-1 sample (from \S{\ref{section:projection-effects}}) and our observational SDSS sample (from \S{\ref{section:sdss}}). Our projected TNG300-1 samples reduce to 241,991 \textit{Ss} and 114,681 \textit{Sp} pairs, and our SDSS samples reduce to 48,055 \textit{Ss} and 45,565 \textit{Sp} pairs.

It is important to note that the control samples for both sets of \textit{Ss} and \textit{Sp} pairs are not limited to only star-forming galaxies as the hosts are. This leads to a disparity in the range of sSFR values between our hosts and controls, since our control samples contain a mix of star-forming and passive galaxies (e.g. 68.7 percent of 3D \textit{Ss} controls and 59.6 percent of 3D \textit{Sp} controls are star-forming), whereas our star-forming host samples only contain star-forming galaxies. We make the conscious choice not to re-match our control samples, and to instead proceed with the original mix of best-matched star-forming and passive control galaxies (as discussed in \S{\ref{section:sample}}). Below, we explain how re-matching our control sample would introduce unintended biases into our data.

Firstly, we consider the physical implications of control matching. We begin with the assumption that we are viewing our host galaxies at an epoch where their own properties have already been affected by the interaction with their closest companion. This must be true (on average) for at least our closest galaxy pairs, or else we would not measure any statistically significant changes in specific star formation rates compared to controls. Knowing that companion interactions do have an effect on star formation rate, but without making any specific assumptions as to how (enhancement or suppression), we can infer that in some cases, host galaxies which were previously passive before interaction may have increased star formation rates as a result of the interactions (and vice versa). If we were to compare our mid-interaction host with its exact pre-interaction counterpart, we may be comparing a star-forming galaxy with a passive one. Thus, in this case, matching our currently-star-forming host with a star-forming control would hide the true effect of the interaction on the star formation rate of the host galaxy.

Secondly, we consider the statistical consequences of re-matching our control galaxies. The key property that we measure in this study is star formation rate. Matching our control sample on the same variable that we are measuring introduces a statistical bias into our data that may artificially minimize differences between hosts and controls. Splitting the sample in this manner would be justified if we had reason to believe we were analyzing two distinct populations of galaxies. However, as discussed above, we must assume that our host population prior to interaction (what we are attempting to emulate with our control sample) contains a mix of star-forming and passive galaxies.

Given the analysis we have already conducted on \textit{Hs} and \textit{Hp} galaxies, we determined that relative trends between \textit{Ss} and \textit{Sp} galaxies are of greater interest to us than specific or absolute values of $Q(\textrm{sSFR})$. To that end, we continue to use our previously best-matched control galaxies regardless of star-forming or passive categorization.

\subsection{Star-forming hosts}
\label{section:sf_hosts}

\begin{figure*}
	\begin{center}
        \subfloat[3D TNG300-1 star-forming hosts]{
            \includegraphics[width=.32\textwidth]{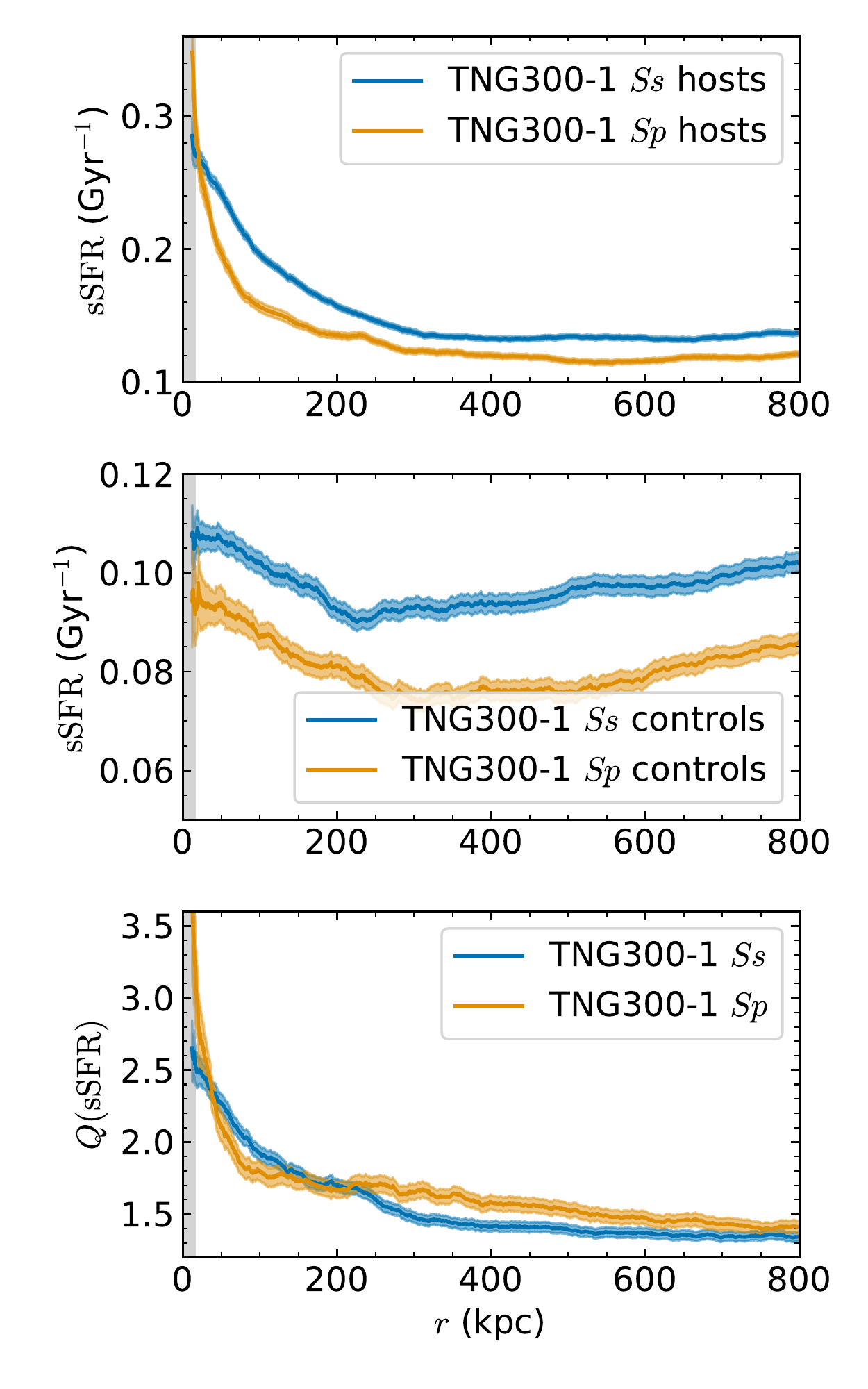}
        }
        \subfloat[Projected TNG300-1 star-forming hosts]{
            \includegraphics[width=.32\textwidth]{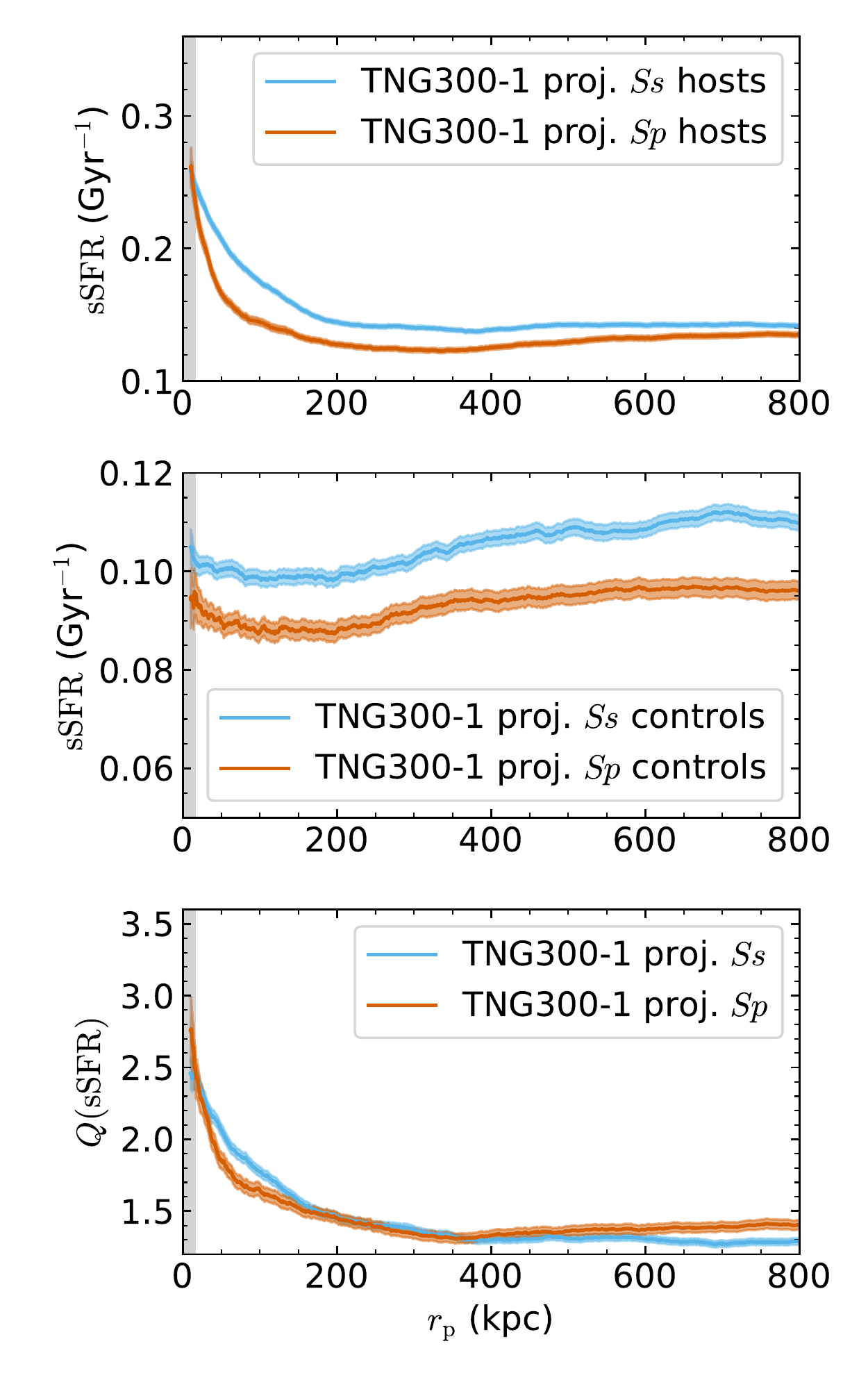}
        }
        \subfloat[SDSS star-forming hosts]{
            \includegraphics[width=.32\textwidth]{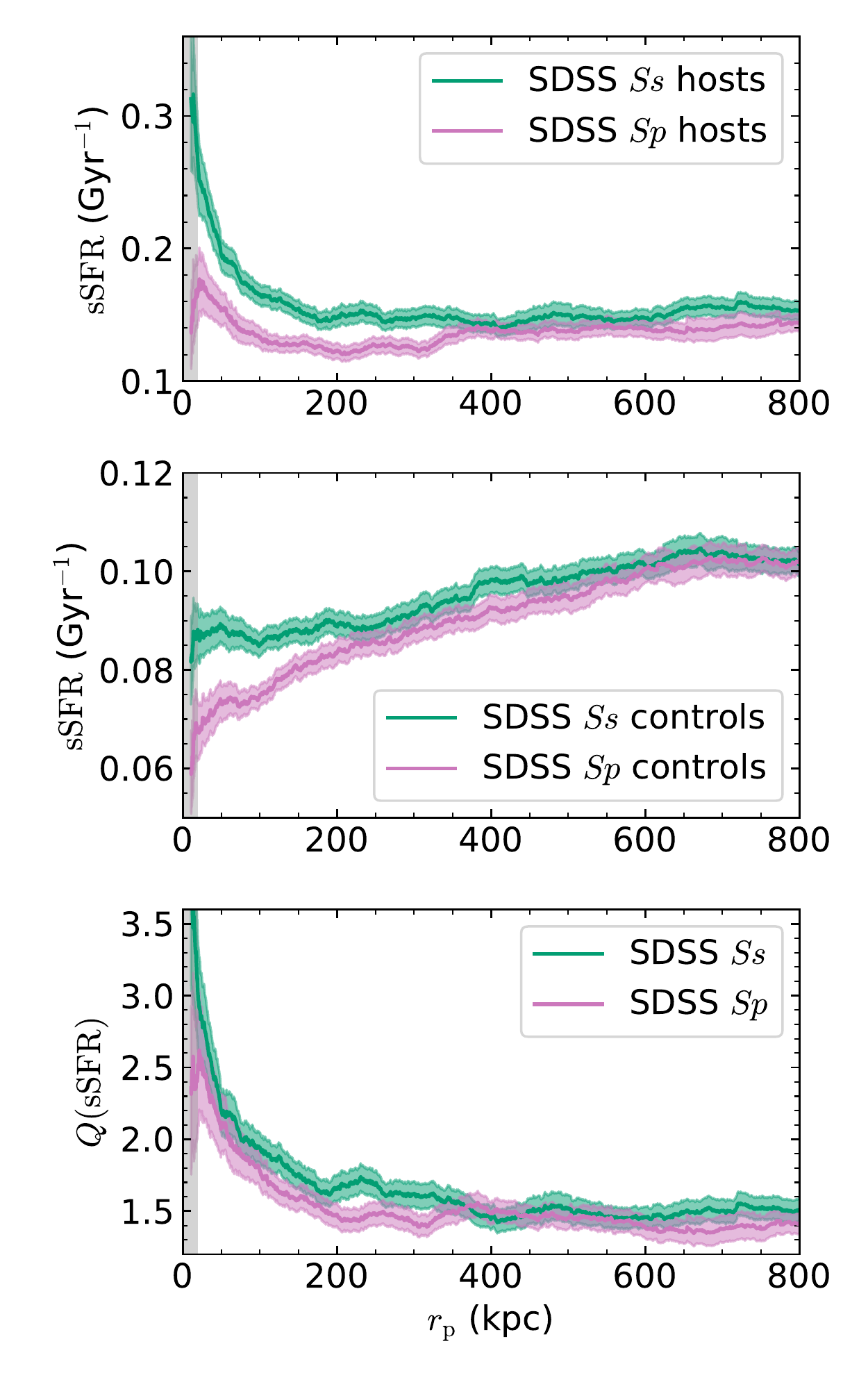}
        }
		\caption{The figures above plot the sSFR enhancement in the 3D TNG300-1 sample (left), the projected TNG300-1 sample (middle), and SDSS sample (right) as a function of galaxy pair separation for \textit{Ss} pairs (blue, light blue, and green) and \textit{Sp} pairs (orange, dark orange, and pink). In the top panel, the mean sSFR of all host galaxies are plotted vs. 3D or projected separation to the host's companion galaxy ($r$ or $r_{\rm p}$ respectively). In the middle panel, the mean sSFR of all control galaxies are plotted vs separation to the host's companion galaxy. In the bottom panel, the interaction-induced sSFR "enhancement" ($Q$) is plotted vs. separation. The grey region represents an approximation of the crowding limit. Shaded regions in all panels represent $2\sigma$ standard error in the mean.}
		\label{fig:host_type}
	\end{center}
\end{figure*}

We plot our results for sSFR vs. separation and $Q\rm(sSFR)$ vs. separation for our 3D TNG300-1 sample ($r$), our our projected TNG300-1 sample ($r_{\rm p}$), and our SDSS ($r_{\rm p}$) in Fig. \ref{fig:host_type}. It is important to note again that our control samples are not matched on the same star-forming/passive categorization as our host galaxies, and that this results in a significant offset between the average control sSFR and the average host sSFR in all categories. We have chosen to plot host and control sSFRs separately, and we advise the reader to pay more attention to the relative shape and trend of $Q\rm(sSFR)$ than the absolute numerical values.

We note that both \textit{Ss} and \textit{Sp} galaxies in both TNG300-1 samples show indications of enhanced sSFR with decreasing 3D separation. Of particular note in Fig. \ref{fig:host_type}(a) is that the trend in \textit{Ss} pairs appears to steadily increase for $r < 300$ kpc, while the trend of \textit{Sp} pairs appears to be much steeper for $r < 75$ kpc. One interpretation of this trend is that passive companions in TNG300-1 must be closer to hosts in order to induce sSFR enhancements compared to star-forming companions. However, galaxy merger simulations \citep{Moreno2019} and observational studies \citep{Patton2013,Pan2019} both suggest that sSFR enhancements at extended $r$ and $r_{\rm p}$ are remnants of starbursts triggered by earlier close pericentre passages \citep{Renaud2022}. With this in mind, the differences in the trends may also indicate that starbursts resulting from close interactions with star-forming companions persist for a longer amount of time than those of passive companions, and can thus be seen at more extended separations as the galaxies move away from each other after a close encounter. sSFR trends in the projected TNG300-1 sample in Fig. \ref{fig:host_type}(b) are similar.

Our SDSS sample agrees with TNG300-1 in that both \textit{Ss} and \textit{Sp} galaxies show evidence of sSFR enhancement. The extent of sSFR enhancements in $r_{\rm p}$ is reduced compared to TNG300-1, which agrees with previous analysis of observational samples in both this work and others (\citetalias{Patton2020} found that the extent of their TNG300-1 enhancements reduced from $r \sim 280$ kpc to $r_{\rm p} \sim 260$ kpc, and enhancements in their sample of star-forming hosts in SDSS extended to $r_{\rm p} \sim 150$ kpc). The biggest difference we note is the disappearance of the steep slope that was present in TNG300-1 \textit{Sp} galaxies. This may be due to meaningful differences between simulations and observations, including perhaps differences in modelled physical processes, simulation resolution, or amount of galaxy area covered by SDSS fibres. We refrain from drawing concrete conclusions from the slope present in TNG300-1 as it is not supported by our observational data.

We use our SDSS sample to conclude that our results disagree with those of \citet{Moon2019}, who found that their equivalent observational \textit{Sp} sample was neither enhanced nor suppressed. While our results show clear evidence of sSFR enhancement, we caution the reader that the methods and samples of both papers differ. As an additional check, we repeated our analysis of SDSS galaxies using a threshold of sSFR = $10^{-2.5}$ Gyr$^{-1}$ to delineate between star-forming and passive galaxies, as was used by \citet{Moon2019}. We find negligible changes in our results between the two sSFR thresholds. However, we suggest the differences may result from the different control matching method used by \citet{Moon2019}, which limits star-forming hosts to being matched only to star-forming controls. Referring to the top and middle panels of Fig. \ref{fig:host_type}(c), we note that had we restricted our control sample to only star-forming galaxies, we may have diluted the strength of our sSFR enhancements in a similar manner to \citet{Moon2019}.

Our analysis of star-forming hosts, particularly that of \textit{Sp} galaxies, presents somewhat contradictory results compared the complete analysis of \textit{Hp} galaxies presented in the rest of this paper. In particular, while \textit{Hp} galaxies as a whole show evidence of sSFR suppression, \textit{Sp} galaxies in both TNG300-1 and SDSS show evidence of sSFR enhancement. We proceed to explore this contradiction by analysing the fraction of passive hosts in our samples in the following section.

\subsection{Passive host fraction}
\label{section:passive_frac}

Despite the intriguing trends uncovered in our analysis of star-forming host galaxies, we elect to avoid direct analysis of passive host galaxies. Measurements of sSFR in galaxies with little-to-no star formation are challenging to measure in both real observations (due to a lack of emission lines) and simulations (due to a dearth of star-forming gas particles). Additionally, it seems ill-advised and futile to attempt to analyse the sSFRs of a group of galaxies which, by definition, have sSFRs close to zero.

\begin{figure}
	\begin{center}
		\includegraphics[width=\columnwidth]{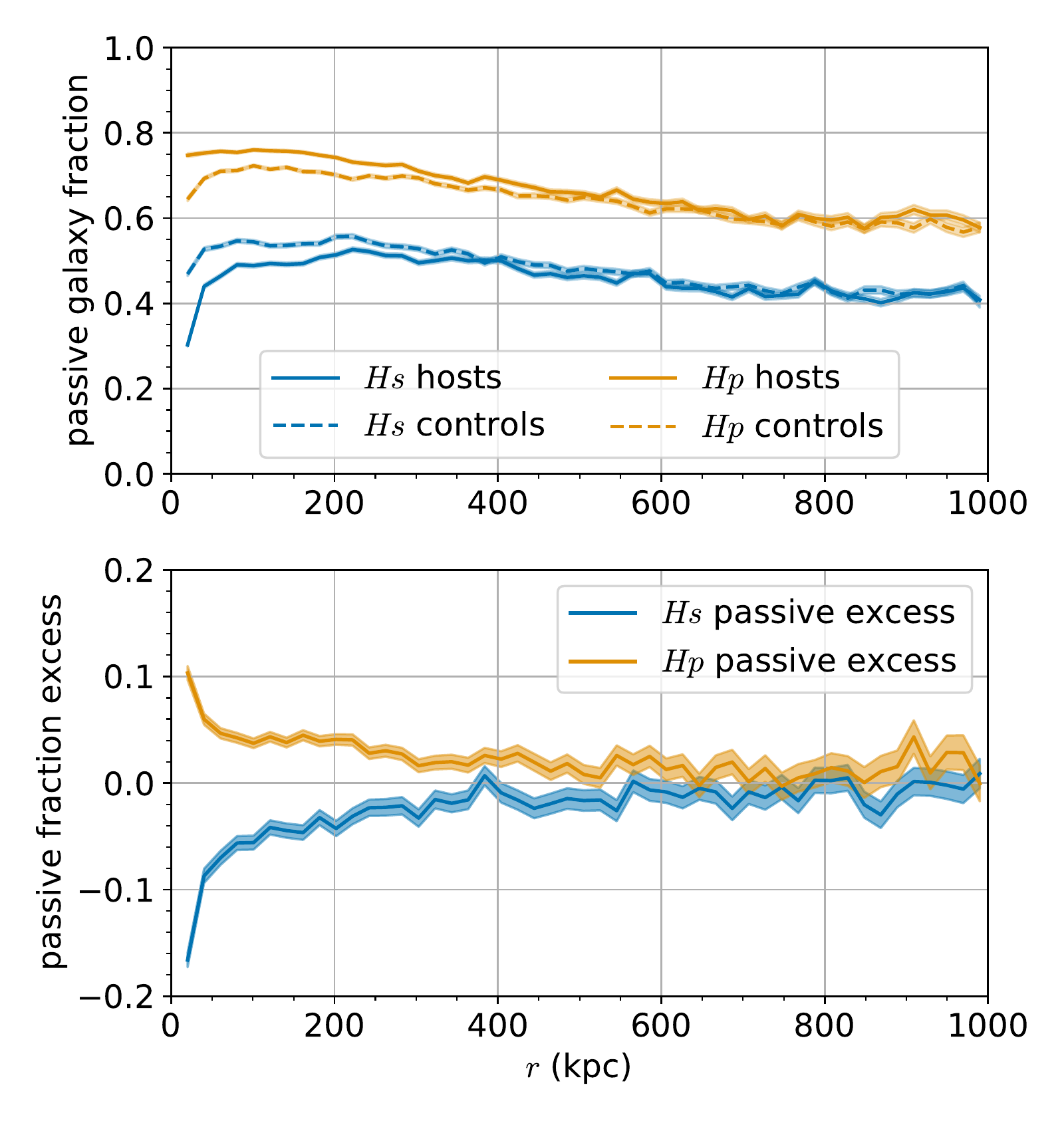}
		\caption{The passive fractions of  \textit{Hs} (blue) and \textit{Hp} (orange) pairs are compared. The upper panel plots the fraction of galaxies that are categorized as passive as a function of 3D separation to closest companion, in fixed $r$ bins. Host galaxies are plotted as solid lines and their corresponding best-matched control galaxies as dashed lines. The bottom panel shows the passive fraction excess for both \textit{Hs} and \textit{Hp} galaxies, which is the difference in the passive fraction of controls and the passive fraction of hosts. A passive fraction excess of 0 represents no change, while a positive number indicates a higher fraction of passive hosts than controls. Shaded regions represent $\sigma$ standard error in the mean.}
		\label{fig:passive_frac}
	\end{center}
\end{figure}

Instead, we infer the strength of companion quenching effects by calculating the fraction of passive hosts as a function of $r$ for both \textit{Hs} and \textit{Hp} pairs in TNG300-1. This is plotted in Fig. \ref{fig:passive_frac}, where we compare the passive fractions of both hosts and controls. In the upper panel, a passive fraction of $0.5$ indicates that half of the galaxies in that bin of $r$ are passive. In the lower panel, we plot the passive fraction excess, which we define as the difference between the passive fraction of host galaxies and the passive fraction of control galaxies. For example, a passive fraction excess of $0.1$ would indicate that an additional 10 percent of host galaxies are passive compared to controls.

We see from Fig. \ref{fig:passive_frac} that while the passive host fraction is anti-correlated with separation at wide separations, this trend reverses at small separations. 
 In particular, both \textit{Hs} and \textit{Hp} passive fractions increase by 10 percent or more from $r \sim 1000$ kpc to $r \sim 200$ kpc, while close star-forming companions are associated with a decrease in passive host fraction when compared to control galaxies. This is consistent with our prior analysis which shows that star-forming companions are associated with sSFR enhancement in host galaxies. Close passive companions, on the other hand, are associated with an increasing fraction of passive hosts compared to controls. While this agrees with most of our prior analysis for \textit{Hp} galaxies, we again point out that it is at odds with the analysis of \textit{Sp} galaxies in \S{\ref{section:sf_hosts}}. If these changes in sSFR are driven by companion type, how is it possible that passive companions can both enhance and suppress the sSFRs of host galaxies? We explore this contradiction, and some possible physics behind it, in \S{\ref{section:discussion}}.

Fig. \ref{fig:passive_frac} also shows the strong impact of galactic conformity in both the \textit{Hs} and \textit{Hp} samples. At large separations ($r > 600$ kpc), 40 percent of \textit{Hs} hosts are passive compared to the 60 percent of \textit{Hp} hosts that are passive. These separations are too large to be accounted for by pair interactions, but they are well within the expected range for environmental effects and galactic conformity. Simply by defining whether the closest companion galaxy 1000 kpc away is star-forming or passive, we have already biased our host galaxy sample. While the trends between hosts and companions appear quite similar at these separations in the upper panel of Fig. \ref{fig:passive_frac}, we note that the bottom panel indicates that host galaxies do experience a small, but still stronger galactic conformity bias than their controls, which can account for the slight $Q\rm(sSFR)$ offsets we have noted in many of our plots.

\textit{Star-forming galaxies with both star-forming and passive companions have enhanced star formation rates on average. However, galaxies with close passive companions are more likely to be passive themselves.}

%%%%%%%%%%%%%%%%%%%%%%%%%%%%%%%%%%%%%%%%%%%%%%%%%%

\section{Discussion}
\label{section:discussion}

\subsection{Potential mechanisms driving sSFR differences}
\label{section:additional_mechs}

In the following sections, we discuss a few physical mechanisms by which galaxy interactions may result in enhancement or suppression of sSFR, and whether or not we find strong indicators of these mechanisms in our own pair samples.

\subsubsection{Hot gas halos}

\citet{Dekel2006} used semi-analytical models to show how galaxies above a critical mass may develop a gaseous halo of shock-heated gas which is stable to gravitational collapse. These hot gas halos then prevent gas within the galaxy from effectively cooling and forming stars, leading to a population of massive, quenched galaxies \citep{Dekel2006,Correa2018}. During a close interaction, a secondary galaxy may pass through the surrounding halo of the primary galaxy. In addition to effects of ram pressure stripping, the hot gas in the halo of the primary galaxy may act like a wall which blocks the secondary galaxy’s access to the cold gas needed to form stars, effectively lowering the star formation rate of the secondary galaxy as its gas supply is cut off and acting as a mechanism for sSFR suppression \citep{Park2009,Gabor2015,Moon2019}. We note that, in contrast to this, \citet{Hwang2015} found that galaxy interactions involving hot gas halos at moderate-wide separations (closest pericentre at $r \sim 100$ kpc) showed evidence for enhanced star formation rates in \textit{N}-body simulations.

\citet{Zinger2020} and \citet{Truong2021} use IllustrisTNG to study the correlation between hot gas, supermassive black holes, and quenching on galactic scales, showing that TNG100-1 and TNG300-1 reasonably simulate important non-star-forming gas mechanics. Were hot gas halos to be a significant mechanism of sSFR suppression in IllustrisTNG pairs, we would expect to find that massive \textit{Hp} companions (pairs with mass ratios of $\mu < 0.5$) have suppressed sSFR at close separations where galaxies might reasonably be expected to pass through one another’s gaseous halos ($r < 50$ kpc). If hot gas halos were a mechanism for sSFR enhancement, as found in \citet{Hwang2015}, we may also expect to find enhanced sSFR in these same pairs at $r > 100$ kpc. In Fig. \ref{fig:mratio}, we have found that in TNG300-1, \textit{Hp} pairs with $\mu < 0.5$ are consistent with no significant enhancement or suppression, and instead the most significant evidence of suppression is present within \textit{Hp} pairs with less massive companions ($\mu > 2$). We can therefore conclude that hot gas halos are not a significant mechanism of sSFR suppression or enhancement in galaxy interactions within IllustrisTNG.

\subsubsection{ISM collisions}

Star-forming galaxies are known to have rich and extended interstellar media (ISM) due to an abundance of gas and dust. While stars can be considered collisionless particles, the ISM cannot. There is evidence that during interactions between two ISM-rich galaxies, the collisions between ISM can cause tidal compressions and increased ISM turbulence, and result in star formation along the area of collision while also increasing star formation efficiency throughout the galaxies \citep{Jog1992,Renaud2014,Renaud2022}. While the location of ISM collision would be beyond the stellar half-mass radius used to measure sSFR in this paper, the effects of star formation efficiency could be present within the measured IllustrisTNG radius and SDSS apertures.

Given the nature of ISM collisions, if they are a significant source of sSFR enhancement in IllustrisTNG, we would expect the resulting SFR enhancement to be most prominent in \textit{Ss} pairs, where sSFR may be higher at more extended separations than that of \textit{Sp} pairs (if \textit{Sp} pairs are enhanced at all). In Fig. \ref{fig:host_type} we find evidence of sSFR enhancement in both \textit{Ss} and \textit{Sp} pairs in TNG300-1. While the average sSFR of \textit{Ss} galaxies appears to be higher than that of \textit{Sp} galaxies, we do still notice significant sSFR enhancement in \textit{Sp} galaxies as well, which cannot be the result of ISM collisions. We posit that ISM collisions are likely at play in some capacity, however the fact that \textit{Sp} galaxies also have enhanced sSFR suggests that the enhancements are driven primarily by some alternative mechanism. ISM collisions are thus unlikely to be a dominant mechanism of star formation enhancement in IllustrisTNG pairs.

\subsubsection{Gas transfer}

Of particular interest when studying differences between star-forming and passive galaxy companions is the concept of gas transfer. Throughout an interaction, a primary galaxy may enhance its own sSFR by siphoning additional star-forming gas from both a gas-rich environment or from a secondary, lower mass galaxy \citep{Hopkins2013,Angles-Alcazar2017}. Gas siphoned from the secondary galaxy is likely taken from the outskirts of the secondary galaxy where the gas is less tightly bound, and transferred to the primary galaxy through along bridges and other tidal features \citep{Sparre2022}. The mechanism of gas transfer would ultimately increase the sSFR of the primary galaxy, while potentially leaving the sSFR of the secondary galaxy unchanged, as we measure sSFR within only one stellar half-mass radius, where the gas is more tightly bound.

For gas transfer to be a significant mechanism behind the sSFR enhancement we see in IllustrisTNG, we would need to see an enhancement in \textit{Hs} pairs, specifically those with less massive companions where $\mu > 2$, regardless of host type. While we do see enhancement in \textit{Hs} pairs across the board, TNG300-1 showed no impact on sSFR enhancement by mass ratio (Fig. \ref{fig:mratio}). Additionally, while we can show that \textit{Ss} pairs have enhanced sSFR, we cannot conclusively say the same for \textit{Hs} pairs with passive hosts. We also see evidence of sSFR enhancement in \textit{Sp} pairs, which cannot be easily explained by gas transfer. For these reasons, it is unlikely that gas transfer is the primary or dominant mechanism behind the sSFR enhancements we see within $R_{1/2}$ IllustrisTNG. However, it is still possible that gas transfer is an effective mechanism for sSFR enhancement in the outskirts of the galaxy which we do not probe.

\subsubsection{Environmentally-driven trends}

While the focus of our study is on the effect of nearby companions on star formation of host galaxies, we have found in multiple scenarios that the effects of galactic conformity are evident in our sample and strong enough to impact our sSFR measurements at large pair separations (see discussion in \S{\ref{section:tng300}} and \S{\ref{section:passive_frac}}). As galactic conformity is an environmental effect, we cannot automatically discount other environmental factors as potential drivers of trends in our results.

To test whether or not environment could be a driving factor in our trends, we explored the possibility that some of the sSFR suppression in \textit{Hp} pairs with $\mu > 2$ may be due to massive, quenched host galaxies which lie at the centres of galaxy clusters. These galaxies may be surrounded by a large number of smaller, passive galaxies which lie below the 10 percent mass threshold to be considered the closest companion. The peak in sSFR suppression in this sample does occur at a similar $r$ value to that of the peak in $N_2$ (the number of companions within 2 Mpc of the host). However, we repeated our sSFR analysis with the highest $N_2$ galaxies removed from the sample and found no significant change in sSFR results. We conclude from this that other environmental effects are well-handled by our control matching process, and not likely to be drivers behind the trends found in this paper.

\subsection{Missing pieces: pre-interaction state and interaction history of the host galaxy}
\label{section:host_history}

As noted in \S{\ref{section:host_type}}, we find conflicting results when analysing sSFR trends of host galaxies with passive companions, such that they are simultaneously associated with enhanced sSFR (see Fig. \ref{fig:host_type}) and suppressed sSFR (see Fig. \ref{fig:t3_Q}, Fig. \ref{fig:sdss}, and Fig. \ref{fig:passive_frac}) of the host galaxy. This tells us that our sSFR trends cannot be due to companion type alone. Our next point of investigation is host type, which makes up the other half of our interacting pair. However, we note that the host sSFR used in this sample is measured at the time of interaction, and thus any host categorization (including the ones made in \S{\ref{section:host_type}}) must be treated as the \textit{mid-interaction host type}---one that has already potentially been altered by the interaction with its companion.

We therefore theorize that we are missing two pieces of key information about the interaction: the pre-interaction state of the host galaxy, and the interaction history. The pre-interaction state of the host galaxy would involve both the star formation rate and gas mass available to the host galaxy prior to significant interaction effects from its companion. This may separate out high sSFR starburst galaxies, steadily star-forming galaxies, passive galaxies that still maintain a small reservoir of gas, and galaxies which are completely quenched of star-forming gas. The interaction history of the galaxy pair can tell us important interaction information, such as the number of previous pericentre passages, minimum separation at pericentre, and time since last pericentre passage. This information is largely absent when we observe galaxy pairs at a single snapshot, which can only tell us current separation and often not tell us whether galaxies are currently approaching or moving away from each other.

As an example of how this information may affect galaxy interactions, we present two theoretical examples of a host galaxy interacting with a passive companion. In the first example, the host galaxy is one with a small reservoir of star-forming gas. Upon first pericentre passage of the companion galaxy, the host galaxy may experience a starburst which briefly enhances its sSFR, before the galaxy burns through its gas. The sSFR of the host would then decrease, and successive pericentre passages of the companion would not enhance the sSFR, resulting in a passive (and suppressed) mid-interaction host galaxy. In our second example, the host galaxy has a large reservoir of star-forming gas, and is capable of sustaining multiple, prolonged starbursts from pericentre passages. The sSFR would continue to be enhanced compared to a non-interacting galaxy and would result in a star-forming mid-interaction host galaxy.

This additional missing information would not only answer questions about interactions with passive companions, but may also provide additional insight into interactions with star-forming companions. Fig. \ref{fig:passive_frac} shows that the fraction of passive host galaxies decreases for close interactions with star-forming companions. This suggests that there may be evidence to support a theory that interactions with star-forming companions may rejuvenate the star formation in some passive hosts, as how some early type galaxies have been found to be rejuvenated by mergers \citep{Rampazzo2007,Fang2012,Jeong2022}. Once again, this is a question that may be answered by knowing the pre-interaction state of the host galaxy.

While galaxy merger simulations are frequently used for this type of analysis, the small range of galaxy types used and the work required to process different interactions limits their ability to fully explore the wide range of galaxy interactions found in the universe. IllustrisTNG and other cosmological simulations are well-equipped to fill in the missing pieces that observational data and studies cannot, as individual galaxies may be followed throughout their history using the snapshot information provided. The galaxy pairs sample used in this paper is not adequate for this type of study, as it was built only to analyse galaxy interactions at specific snapshots, similar to that of observational data.  However, in a forthcoming study (Patton et al. in prep), we will track the orbital histories of TNG100-1 galaxy pairs, thereby creating samples that can be used to track changes in galaxy properties as a function of encounter stage.  This approach will allow us to distinguish between galaxy properties that were in place before a recent close encounter vs. galaxy properties that have changed as a result of a close encounter. For example, Faria et al. (in prep) will examine how the star-forming properties of IllustrisTNG galaxies change during and after close encounters. These studies will further our understanding of galaxy interactions within a realistically simulated cosmological environment.

%%%%%%%%%%%%%%%%%%%%%%%%%%%%%%%%%%%%%%%%%%%%%%%%%%

\section{Conclusions}
\label{section:conclusions}

Using closest companion galaxy samples developed by \citet{Patton2020} from the TNG100-1 and TNG300-1 runs of the IllustrisTNG cosmological simulations \citep{Nelson2019a}, we have studied the impact of close companion type on the specific star formation rate (sSFR) enhancement of host galaxies. We have divided our sample into host galaxies whose closest companion is star-forming (\textit{Hs}) or passive/quiescent (\textit{Hp}), and we have studied the ratio of sSFR enhancement ($Q(\rm{sSFR})$) of the hosts compared to controls as a function of 3D separation ($r$) from the closest companion.

In addition to our main sample, we also studied \textit{Hs} and \textit{Hp} enhancements split up into three sub-categories based on the stellar mass ratio of the host galaxy to that of the companion. We developed a "twinned" sample by matching \textit{Hs} and \textit{Hp} galaxies to each other based on host and companion stellar mass, creating two samples with equivalent mass distributions. Our results from TNG100-1 and TNG300-1 were compared to observational SDSS results by generating equivalent samples in projected space. Finally, we analyzed a reduced sample of only star-forming hosts (\textit{Ss} and \textit{Sp}) and compared the fraction of passive hosts between \textit{Hs} and \textit{Hp} galaxies as a function of $r$.

Our main conclusions are as follows:

\begin{enumerate}
    \item \textit{Hs} galaxies have enhanced sSFR in IllustrisTNG, to a maximum enhancement factor of $Q=2.9\pm0.3$ and extent of $r \sim 300$ kpc in TNG100-1, and maximum enhancement of $Q=2.27\pm0.06$ and extent of $r \sim 350$ kpc in TNG300-1. (Fig. \ref{fig:t1_Q} for TNG100-1 and Fig. \ref{fig:t3_Q} for TNG300-1)
    \item Suppressed sSFR is seen in \textit{Hp} galaxies at moderate ($\sim 200$ kpc) 3D pair separations. This suppression is not driven by \textit{Sp} pairs, but possibly due to an increased fraction of passive host galaxies with decreasing pair separation. (Fig. \ref{fig:t3_Q}, Fig. \ref{fig:host_type}, and Fig. \ref{fig:passive_frac})
    \item The sSFR of \textit{Hp} galaxies is strongly affected by the mass ratio of host galaxy to companion such that more massive hosts with less massive companions will experience more significant changes to sSFR compared to controls. The sSFR of \textit{Hs} galaxies is generally unaffected by the mass ratio of host galaxy to companion. (Fig. \ref{fig:mratio})
    \item A twinned sample of mass-matched \textit{Hs} and \textit{Hp} pairs finds agreement with our original non-twinned sample, meaning that differences between \textit{Hs} and \textit{Hp} pairs are not driven by intrinsic mass differences between star-forming and passive galaxies. (Fig. \ref{fig:mass-match_Q})
    \item sSFR trends in projected pairs from SDSS are broadly consistent with those in projected pairs from TNG100-1 and TNG300-1. (Fig. \ref{fig:sdss})
    \item Results from our observational sample of SDSS galaxies disagree with the results found by \citet{Moon2019}, in that we do find sSFR enhancements in \textit{Sp} galaxies. (Fig. \ref{fig:host_type})
    \item The extent of our analysis allows us to conclude that commonly explored mechanisms for sSFR enhancement/suppression, such as hot gas halos, ISM collisions, gas transfer, and environment are not sufficient to fully explain all the trends found in this paper. We theorize that the pre-interaction state and interaction history of the host galaxy may play a significant role in determining sSFR enhancements in interactions with companions. (See \S{\ref{section:discussion}})
\end{enumerate}

%%%%%%%%%%%%%%%%%%%%%%%%%%%%%%%%%%%%%%%%%%%%%%%%%%

%%%%%%%%%%%%%%%%%%%%%%%%%%%%%%%%%%%%%%%%%%%%%%%%%%

\section*{Acknowledgements}

We thank all members of the IllustrisTNG collaboration for making their data available to the community. DRP and SLE gratefully acknowledge NSERC of Canada for Discovery Grants which helped to fund this research.  WB was supported in part by an NSERC graduate scholarship, an Ontario graduate scholarship, and an NSERC Undergraduate Student Research Award.

Funding for the SDSS and SDSS-II has been provided by the Alfred P. Sloan Foundation, the Participating Institutions, the National Science Foundation, the U.S. Department of Energy, the National Aeronautics and Space Administration, the Japanese Monbukagakusho, the Max Planck Society, and the Higher Education Funding Council for England. The SDSS Web Site is \url{http://www.sdss.org/}.

The SDSS is managed by the Astrophysical Research Consortium for the Participating Institutions. The Participating Institutions are the American Museum of Natural History, Astrophysical Institute Potsdam, University of Basel, University of Cambridge, Case Western Reserve University, University of Chicago, Drexel University, Fermilab, the Institute for Advanced Study, the Japan Participation Group, Johns Hopkins University, the Joint Institute for Nuclear Astrophysics, the Kavli Institute for Particle Astrophysics and Cosmology, the Korean Scientist Group, the Chinese Academy of Sciences (LAMOST), Los Alamos National Laboratory, the Max-Planck-Institute for Astronomy (MPIA), the Max-Planck-Institute for Astrophysics (MPA), New Mexico State University, Ohio State University, University of Pittsburgh, University of Portsmouth, Princeton University, the United States Naval Observatory, and the University of Washington.

%%%%%%%%%%%%%%%%%%%%%%%%%%%%%%%%%%%%%%%%%%%%%%%%%%

\section*{Data Availability}

The IllustrisTNG data used in this work is publicly available at \url{http://www.tng-project.org}.

SDSS DR7 data is publicly available at \url{https://www.sdss.org/}.  SDSS stellar mass estimates were published in \citet{Mendel2014} and are publicly available at \url{http://www.astro.uvic.ca/~tmendel/stellar_mass/}.  SDSS star formation rates from \citet{Brinchmann2004} are publicly accessible in the MPA/JHU SDSS DR7 catalogs: \url{https://wwwmpa.mpa-garching.mpg.de/SDSS/DR7/}.

%%%%%%%%%%%%%%%%%%%%%%%%%%%%%%%%%%%%%%%%%%%%%%%%%%



\bibliographystyle{mnras}
\bibliography{References}

%\begin{thebibliography}{99}
%\end{thebibliography}

%%%%%%%%%%%%%%%%%%%%%%%%%%%%%%%%%%%%%%%%%%%%%%%%%%

%%%%%%%%%%%%%%%%% APPENDICES %%%%%%%%%%%%%%%%%%%%%

%%%%%%%%%%%%%%%%%%%%%%%%%%%%%%%%%%%%%%%%%%%%%%%%%%

% Don't change these lines
\bsp	% typesetting comment
\label{lastpage}
\end{document}